\documentclass[pteplogo]{ptephy_v1}

\usepackage[T1]{fontenc}
\usepackage{mathtools}
\allowdisplaybreaks
\usepackage{slashed}
\usepackage{MnSymbol}

\begin{document} 

\title{Angular Momentum Inheritance from the Schwinger Effect in (Chromo)electromagnetic Fields}

\author{Patrick Copinger}
\affil{Institute of Physics, Academia Sinica, Taipei 11529, Taiwan \email{copinger0@gate.sinica.edu.tw}}
\author[2]{Yoshimasa Hidaka}
\affil{KEK Theory Center, Tsukuba 305-0801, Japan, Graduate University for Advanced Studies (Sokendai), Tsukuba 305-0801, Japan, RIKEN iTHEMS, RIKEN, Wako 351-0198, Japan \email{hidaka@post.kek.jp}}


\begin{abstract}
The angular momentum of fermion pairs generated by the Schwinger effect is studied in homogeneous (chromo)electromagnetic fields, mimicking the early stages of a heavy-ion collision. It is demonstrated that the angular momentum density of produced pairs is proportional to that of the background fields. This is argued both heuristically in a virtual breaking condensate model by evaluating Wong's equations, and out-of-equilibrium to one-loop using the in-in formalism.
\end{abstract}

\maketitle
\flushbottom

\section{Introduction}
\label{sec:intro}

Off-central heavy-ion collisions (HIC) of two nuclei produce enormous angular momenta and are thought to give rise to the ``most vortical fluid'' in the world~\cite{Liang:2004ph,PhysRevC.77.044902,PhysRevC.76.044901}. A by-product of this is thought to be the spin polarization of the $\Lambda$ and $\bar{\Lambda}$ hyperons measured by the STAR Collaboration at RHIC~\cite{STAR:2017ckg,PhysRevC.98.014910}. Several studies dissecting the transference of relativistic vorticity to polarization exist, and include but are not limited to those on: the Einstein-de Haas and Barnett effects~\cite{1915DPhyG..17..152E,RevModPhys.7.129,PhysRevA.99.032105}, hydrodynamic models~\cite{Karpenko:2016jyx,PhysRevC.96.054908}, the role of spin-orbit coupling~\cite{PhysRevC.76.044901,Liang:2004ph}, and quantum kinetic theory based on a Wigner formalism~\cite{PhysRevLett.109.232301,PhysRevC.94.024904,Pang_2017}. Moreover, the global/local polarization problem has renewed interest in the topic of angular momentum transport in HIC and their frameworks for facilitation, namely quantum electrodynamics (QED) and quantum chromodynamics (QCD). A phenomenon important in QED and QCD, but that has received little attention to its quantum transport qualities is the Schwinger effect.

The Schwinger effect is a nonperturbative process that predicts the quantum field theory vacuum is unstable against the production of particle-antiparticle pairs in the presence of a strong electric field~\cite{sauter,Heisenberg:1935qt,PhysRev.82.664}. A feature of the Schwinger effect is an inheritance of the properties of the field onto the produced particles. Not only may the particles acquire energy and momentum (depending on the makeup of background field), but also parity violating characteristics. Namely, under $\mathcal{CP}$-odd background fields, such as parallel electric and magnetic fields, through the Schwinger effect, produced particles are $\mathcal{CP}$-odd, whose relationship  is described by the chiral anomaly~\cite{PhysRevLett.104.212001,PhysRevD.86.085029,Tanji:2010eu,PhysRevLett.121.261602}. Then it is intuitive and important to address whether the Schwinger effect can provide the means to transport angular momentum from field to constituents. Note that Schwinger effect should furnish angular momentum has been assumed in refs.~\cite{PhysRevD.78.036008,PhysRevD.100.045015}.

While the Schwinger effect in quantum electrodynamics (QED) is strongly suppressed in, e.g., experimental setups at high-power laser facilities~\cite{lasers} (where it still remains unseen), for the strong fields in heavy-ion collisions (HIC), the Schwinger effect is thought to underlie chromoelectric flux-tube breaking leading to hadronization~\cite{PhysRevD.20.179}. In the early stages of an HIC, a dense gluonic state forms called the glasma~\cite{KHARZEEV2002298,LAPPI2006200}, where such flux-tubes are thought to be present. The Schwinger effect in non-Abelian fields has been explored in ref.~\cite{TANJI20102018}, the effect for gluons in ref.~\cite{GYULASSY1985157}, Nielsen-Olesen unstable modes in ref.~\cite{TANJI2012117}, and also its role in topological background fields in ref.~\cite{PhysRevD.103.036004}.

We explore the inheritance of orbital and spin angular momentum via the Schwinger effect from both a heuristic standpoint and using an out-of-equilibrium in-in framework. Specifically for the former, we treat a virtual breaking condensate model in which generated pairs evolve classically according to Wong's equations~\cite{Wong:1970fu,PhysRevD.17.3247,PhysRevD.15.2308}. This approach is known to agree with calculations up to one-loop, e.g., the axial-vector and vector currents associated with pair production~\cite{doi:10.1142/S0217751X2030015X}, and is physically transparent. We also analyze an out-of-equilibrium full quantum in-in construction~\cite{etde_6972673}. Background fields are taken as immutable, in which backreaction effects may be ignored.

We first introduce the (chromo)electromagnetic background field setup that is motivated by an HIC in section~\ref{sec:hic}. Next we derive and then evaluate Wong's equations leading to a heuristic picture of pair production in section~\ref{sec:wong}, and then confirm a similar angular momentum quantity to one-loop and out-of-equilibrium in section~\ref{sec:nonequilibrium}. Conclusions are last presented in section~\ref{sec:conclusions}.

We use natural units such that $c=\hbar=1$, and we work entirely in Minkowski spacetime with $g_{\mu\nu}=\textrm{diag}(+,-,-,-)$. Our covariant derivative is defined as $\mathcal{D}_{\mu}=\partial_{\mu}+ieA_{\mu}(x)+ig\mathcal{A}_{\mu}(x)$; for the SU$(2)$ fields we use the fundamental representation such that $\mathcal{A}_{\mu}=\mathcal{A}_{\mu}^{a}T^{a}$, with $T^{a}=(1/2)\sigma^{a}$ being the usual Pauli matrices. The spin tensor reads $\sigma_{\mu\nu}=\frac{i}{2}[\gamma_{\mu},\gamma_{\nu}]$. In the coincidence limit we take $S(x,x)=\lim_{\epsilon\rightarrow 0}(1/2)[S(x,x+\epsilon)+S(x+\epsilon,x)]$, which will also define the Heaviside theta function to be $\theta(0)=1/2$, about the origin where present. Finally, throughout this paper where appropriate we make use of a compact matrix notation in Lorentz indices, i.e., $F_{\;\;\nu}^{\mu} \eqqcolon F$, and $x^\mu \eqqcolon x$ is a column vector with $x_\mu  \eqqcolon x^T$.

\section{HIC Event Averaged Angular Momentum}
\label{sec:hic}

Our task is to explore the simplest possible theoretical setup in which a background field both possesses a net angular momentum and resembles the early stages of an HIC. We find this can be had utilizing fields diagonal in color as are thought present in the chromoelectric flux tubes~\cite{KHARZEEV2002298,LAPPI2006200}, whereby upon averaging over a number of collision events, a net angular momentum would be finite.  To accomplish this it is sufficient that we treat homogeneous and diagonal in color--or rather Abelianized--non-Abelian fields, namely with
\begin{equation}\label{eq:abelianized}
\mathcal{A}_\mu\propto \sigma_3\,,
\end{equation}
for our case of SU$(2)$. To discuss the event-by-event averaged profile, let us first define the homogeneous background electric and magnetic fields and their properties for the Abelian and SU$(2)$ non-Abelian fields with isospin $I$ respectively as
\begin{alignat}{3}
    E^i & =F^{i0}\,, && B^i&&=\widetilde{F}^{i0}\,,\\
    \mathcal{E}^i & =\mathrm{tr}_c[IG^{i0}]\,,\quad &&\mathcal{B}^i &&=\mathrm{tr}_c[I\widetilde{G}^{i0}]\,,
    \label{eq:homo_fields}
\end{alignat}
for $\widetilde{F}^{\mu\nu}=\frac{1}{2}\epsilon^{\mu\nu\alpha\beta}F_{\alpha\beta}$, $\widetilde{G}^{\mu\nu}=\frac{1}{2}\epsilon^{\mu\nu\alpha\beta}G_{\alpha\beta}$, for field strength given by
$F_{\mu\nu}=\partial_\mu A_\nu-\partial_\nu A_\mu$,  and $G_{\mu\nu}=\partial_{\mu}\mathcal{A_{\nu}}-\partial_{\nu}\mathcal{A_{\mu}}+ig[\mathcal{A}_{\mu},\mathcal{A}_{\nu}]$. 
Here, $\epsilon^{\mu\nu\alpha\beta}$ is the totally antisymmetric tensor with $\epsilon^{0123}=+1$.  The isospin matrix, $I\in$ SU$(2)/$U$(1)$, characterizes a coupling of a particle with color to its background non-Abelian field (The definition of this matrix will be given in eq.~\eqref{eq:isospin}).  $\mathrm{tr}_c$ denotes a trace with respect to color indices. The isospin and field strength change under a  color gauge transformation, $U\in$ SU$(2)$, as $IG^{\mu\nu}\rightarrow U^\dagger IU U^\dagger G^{\mu\nu} U$, and therefore the descriptions given in eq.~\eqref{eq:homo_fields} are gauge-invariant.  Moreover, due to the Abelianized field assumption in eq.~\eqref{eq:abelianized}, we will find that the isospin too takes on a diagonal in color representation with $I\propto \sigma_3$. Therefore,  the field strength of our system may be taken as two independent homogeneous Abelian stengths in superposition:
\begin{equation}
    \mathcal{F}_{\mu\nu}\coloneqq eF_{\mu\nu}+g \mathrm{tr}_c [IG_{\mu\nu}]\,.
    \label{eq:F_tot}
\end{equation}
Then one may write for the Lorentz invariants of the combined field strength tensor
\begin{align}
    I_{\tilde{F}F}&=-\frac{1}{4}\widetilde{\mathcal{F}}_{\mu\nu}\mathcal{F}^{\mu\nu}=(g\boldsymbol{\mathcal{E}}+e\boldsymbol{E})\cdot(g\boldsymbol{\mathcal{B}}+e\boldsymbol{B})\,,\\
    I_{FF} & =\frac{1}{2}\mathcal{F}_{\mu\nu}\mathcal{F}^{\mu\nu}=(g\boldsymbol{\mathcal{B}}+e\boldsymbol{B})^2-
    (g\boldsymbol{\mathcal{E}}+e\boldsymbol{E})^2\,.
    \label{eq:Lorentz_invariants}
\end{align}
For the symmetric stress tensor of Abelian fields,
$\Theta^{\mu\nu}={F}^{\mu\alpha}{F}_{\alpha}^{\,\,\nu}+\frac{1}{4}g^{\mu\nu}{F}_{\alpha\beta}{F}^{\alpha\beta}$, e.g.,~\cite{Jackson:1998nia}, one can find the relativistic angular momentum density as  $\Theta^{\mu\nu}x^{\sigma}-\Theta^{\mu\sigma}x^{\nu}$. An analogous symmetric stress tensor and relativistic angular momentum density can be found for non-Abelian particles by summing over color indices,
$2\mathrm{tr}_c[{G}^{\mu\alpha}{G}_{\alpha}^{\,\,\nu}]+\frac{1}{2}g^{\mu\nu}\mathrm{tr}_c[{G}_{\alpha\beta}{G}^{\alpha\beta}]$.
It is however convenient to cast the gluonic relativistic angular momentum density with isospin, of which the spatial components read
\begin{equation}
    \boldsymbol{l}_{\mathcal{F}}=\boldsymbol{x}\times(\boldsymbol{\mathcal{E}}\times \boldsymbol{\mathcal{B}})\,.
    \label{eq:F_angular_density}
\end{equation}
We note that $\boldsymbol{l}_{\mathcal{F}}$ is different from the ordinary angular momentum of gluons because $\boldsymbol{\mathcal{E}}$ and $\boldsymbol{\mathcal{B}}$ contain the isospin matrix. If one takes the average over the isospin matrix with uniform weight, it reduces to the ordinary one.

One may define a total spatial angular momentum density which includes Abelian fields with the addition of $\boldsymbol{x}\times(\boldsymbol{E}\times \boldsymbol{B})$ to the above, however for the background fields discussed below, such a contribution will vanish.

\begin{figure}
    \centering
    \includegraphics[scale=0.5]{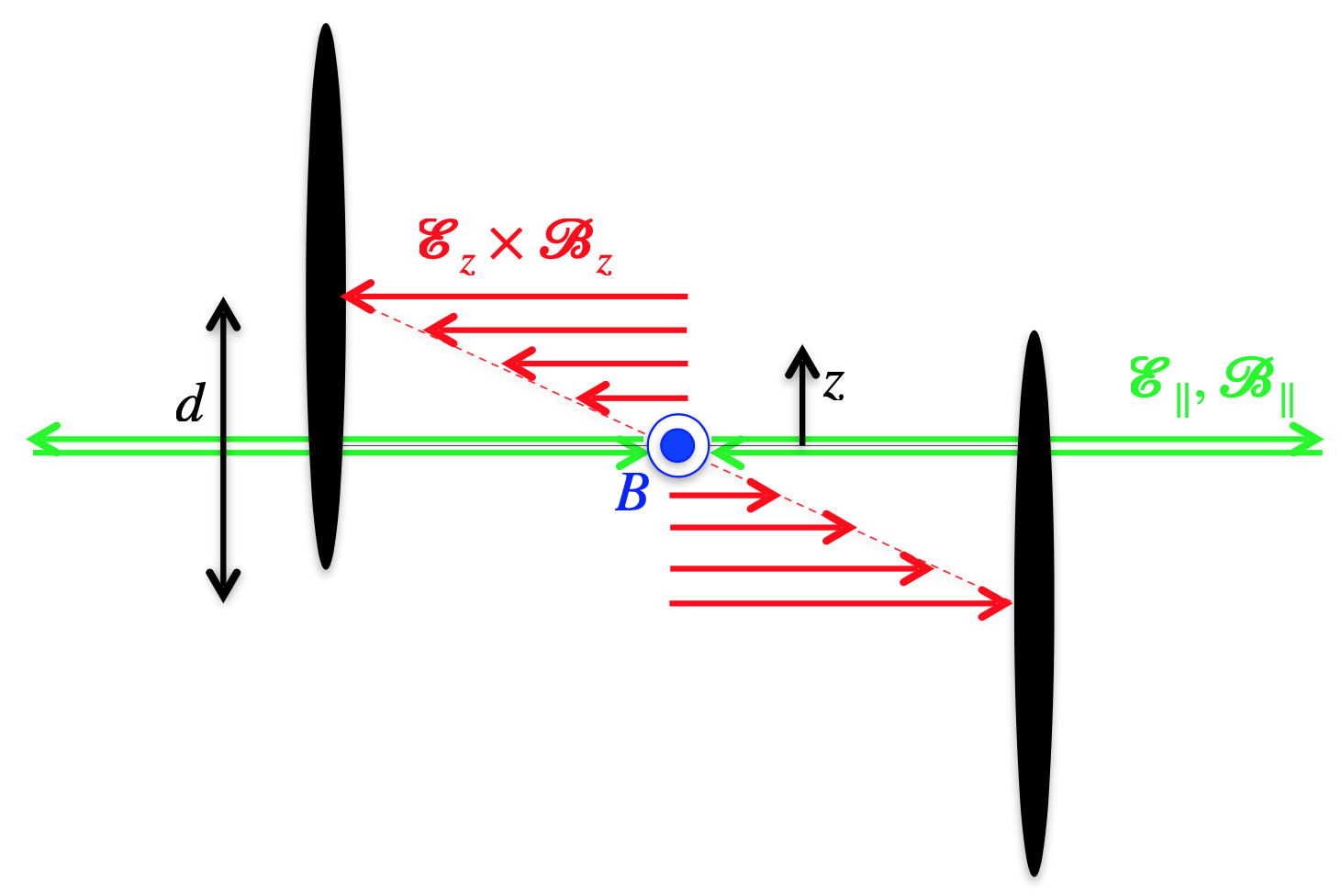}
    \caption{U$(1)$ and SU$(2)$ background field theoretical model. The simplest case motivated by early stages of an HIC event, with offset parameter, $d/2$, and gluon momentum, $\boldsymbol{\mathcal{E}}\times\boldsymbol{\mathcal{B}}$, is considered. Fields are treated as homogeneous. Inhomogeneity as depicted in the figure results from an averaging over events. For a given event, we take eq.~\eqref{eq:EcrossB_setup} for some $z\in[-d/2,d/2]$ and we also take eq.~\eqref{eq:parallel_configs}, where put together, eq.~\eqref{eq:hic_field_decom}, $\boldsymbol{\mathcal{E}}$ and $\boldsymbol{\mathcal{B}}$ may be treated as homogeneous. A strong out-of-plane Abelian magnetic field, $\boldsymbol{B}=B\hat{\boldsymbol{x}}_{\perp}$, can also be seen.}
    \label{fig:hic}
\end{figure}

Then for two off-central oncoming nuclei, with offset parameter, $d/2$, that have both gluon momentum and a parity violating component in the lab frame we confine our attention to the setup given in figure~\ref{fig:hic}. For a given event the components of the electric and magnetic chromo fields are decomposed as
\begin{equation}
    \boldsymbol{\mathcal{E}}= \boldsymbol{\mathcal{E}}_z +\boldsymbol{\mathcal{E}}_\parallel\,,\quad
    \boldsymbol{\mathcal{B}}= \boldsymbol{\mathcal{B}}_z +\boldsymbol{\mathcal{B}}_\parallel\,.
    \label{eq:hic_field_decom}
\end{equation}
On each HIC event a net gluon momentum in the collision direction is taken such that the total momentum over many events is zero, however, the total angular momentum is finite. 
We analyze a small space-time volume, $\mathcal{VT}$, over which Schwinger pair production is assumed occur, and we treat within the small volume homogeneous fields. Then we evaluate the effects of Schwinger produced pairs for a given event and then average over all events. 
We treat the glasma flux-tube model of pair production (see~\cite{PhysRevLett.104.212001} and its references therein for its usage under homogenous and Abelian projected fields).  The available space-time volume for pair production in a glasma flux tube is small, however it is sufficient.  For the model we assume $g\mathcal{E}\sim g\mathcal{B}\sim Q_{s}^{2}\sim 1$ GeV$^{2}$  and that is also homogeneous over a spatial scale of $\Delta_{T}\sim Q_{s}^{-1}$, where the saturation scale is given by $Q_s$. We assume the quark mass is small in comparison to a transverse momentum, which we assume is $p_{T}\sim\Delta_{T}^{-1}\sim Q_{s}$ corresponding to the transverse size in a flux tube. The room available for pair production is approximately given by the exponental suppression factor as a function of transverse momentum, i.e., $p_{T}/g\mathcal{E}\sim Q_{s}^{-1}$, which we can see is of the same order as the spatial scale of $\Delta_{T}$; in this way pair production would be applicable in the given small space-time volume. Nevertheless, in a real collision the fields are inhomogeneous, and we use the idealized approximation of homogeneous fields for analytic tractability. We can approximate this by assuming the idealized situation of $Q_{s}\ll\Delta_{T}^{-1},$ which enables the clear relation of angular momentum we will put forth.
We also assume that the fields in the flux tubes have a component that may be oriented so that $I_{\tilde{F}F}$ scales as the saturation scale, $Q_s^4$, but when averaged over are globally zero. A background Abelian magnetic field pointing in the out-of-plane direction, $\hat{\boldsymbol{x}}_{\perp}$, is also assumed; $\boldsymbol{B}=B\hat{\boldsymbol{x}}_{\perp}$,
while an electric field is assumed to vanish.  Although electromagnetic magnetic field may reach as high as $10^4$ MeV~ in HIC~\cite{KHARZEEV2008227,magneticfield}, we assume a scenario in which the electromagnetic magnetic field is $10-100$ MeV to better study the transference of angular momentum from the chromo fields.
We assume for the superposed chromoelectromagnetic field in eq.~\eqref{eq:hic_field_decom}, that the component in the beam direction is entirely sourced by $\boldsymbol{\mathcal{E}}_\parallel$ and $\boldsymbol{\mathcal{B}}_\parallel$, therefore we have that $\boldsymbol{\mathcal{E}}_\parallel\cdot \boldsymbol{\mathcal{E}}_z=  \boldsymbol{\mathcal{B}}_\parallel \cdot \boldsymbol{\mathcal{E}}_z= \boldsymbol{\mathcal{E}}_\parallel\cdot \boldsymbol{\mathcal{B}}_z=\boldsymbol{\mathcal{B}}_\parallel\cdot \boldsymbol{\mathcal{B}}_z=0$. We also assume that all $\mathcal{CP}$ violation comes from $\boldsymbol{\mathcal{E}}_\parallel$ and $\boldsymbol{\mathcal{B}}_\parallel$ and therefore $\boldsymbol{\mathcal{E}}_z$ and $\boldsymbol{\mathcal{B}}_z$ are perpendicular.  Finally,  in assuming that the strength of both are similar, i.e., $|\boldsymbol{\mathcal{E}}_z| \sim |\boldsymbol{\mathcal{B}}_z|$, we find that the pair production be sourced by $\boldsymbol{\mathcal{E}}_\parallel$ and $\boldsymbol{\mathcal{B}}_\parallel$ since then $I_{\tilde{F}F}\approx g^2\boldsymbol{\mathcal{E}}_{\parallel}\cdot\boldsymbol{\mathcal{B}}_{\parallel}$ and $I_{FF}\approx g^{2}(\boldsymbol{\mathcal{B}}_{\parallel}\cdot\boldsymbol{\mathcal{B}}_{\parallel}-\boldsymbol{\mathcal{E}}_{\parallel}\cdot\boldsymbol{\mathcal{E}}_{\parallel})$.  

Let us first examine the gluon momentum part of eq.~\eqref{eq:hic_field_decom}; we take the momentum to be
\begin{equation}
    \boldsymbol{\mathcal{E}}_z\times\boldsymbol{\mathcal{B}}_z=\mathcal{E}_z^{\text{max}}\mathcal{B}_z^\text{max}\frac{2z}{d}\hat{\boldsymbol{x}}_{\parallel}\,,
    \label{eq:EcrossB_setup}
\end{equation}
where for each event $z$ is taken as an external parameter, and the fields, $\boldsymbol{\mathcal{E}}_z$ and $\boldsymbol{\mathcal{B}}_z$, may then be treated as homogeneous.  $z\in[-d/2,d/2]$ is oriented perpendicular to the Abelian magnetic field and the gluon momentum as shown in figure~\ref{fig:hic}.  $\hat{\boldsymbol{x}}_{\parallel}$ denotes the direction parallel to the beam axis.  With the exception of the fields described above,  any configuration of $\boldsymbol{\mathcal{E}}_z$ or $\boldsymbol{\mathcal{B}}_z$ leading to eq.~\eqref{eq:EcrossB_setup} is accepted.

Next, let us look at the parity violating part of eq.~\eqref{eq:hic_field_decom}. Here we assume that $\boldsymbol{\mathcal{E}}_\parallel$ and $\boldsymbol{\mathcal{B}}_\parallel$ may be oriented parallel or anti-parallel to the beam axis so that $I_{\tilde{F}F} \approx \pm g^2\mathcal{E}_\parallel \mathcal{B}_\parallel$, where
\begin{equation}
    g^2\mathcal{E}_\parallel \mathcal{B}_\parallel\sim Q_s^4\,.
\end{equation}
To achieve this we take specifically 
\begin{equation}
    \boldsymbol{\mathcal{E}}_\parallel=\pm \mathcal{E}_\parallel \hat{\boldsymbol{x}}_{\parallel}\,,\quad
    \boldsymbol{\mathcal{B}}_\parallel=\pm \mathcal{B}_\parallel \hat{\boldsymbol{x}}_{\parallel}\,.
    \label{eq:parallel_configs}
\end{equation}

We take for an event the quantum expectation value of some observable (or background field), $ o_z$, where for the given event $z$ is treated as a fixed parameter. Then we may average over all events such that
\begin{equation}
    \llangle o \rrangle \coloneqq \frac{1}{4}\sum_{\pm \mathcal{E}_\parallel,\pm \mathcal{B}_\parallel}\frac{1}{d}\int^{d/2}_{-d/2}dz\, o_z \,.
    \label{eq:hic_event}
\end{equation}
For example, one may readily find for the average momentum of the fields as $\llangle \boldsymbol{\mathcal{E}}_z \times \boldsymbol{\mathcal{B}}_z \rrangle=0$. However, the average chromoelectromagnetic angular momentum of the system as viewed from the center of the collision using eq.~\eqref{eq:F_angular_density} can be found as
\begin{equation}
    \llangle\boldsymbol{l}_{\mathcal{F}}\rrangle = \mathcal{E}_z^\text{max}\mathcal{B}_z^\text{max}\frac{d}{6} \hat{\boldsymbol{x}}_{\perp}\,,
    \label{eq:F_total_ang_mom}
\end{equation}
resulting in a net angular momentum of the fields over all events. Note that here a contribution to the angular momentum from the Abelian magnetic field vanishes once averaged over in $z$. Last, we also have for a global parity violation that 
\begin{equation}
    \llangle I_{\tilde{F}F} \rrangle \approx g^2\llangle \boldsymbol{\mathcal{E}}_\parallel \cdot \boldsymbol{\mathcal{B}}_\parallel \rrangle = 0\,.
\end{equation}

    Let us point out that realistic collisions differ from the simplified model, although we analyze the above scenario that mimics certain aspects of an HIC. We restrict our study to those of homogeneous fields, and this is for physical opacity as well as benefitting from exactly solvable setups; however, the fields in a HIC are inhomogeneous~\cite{Tuchin:2013ie}. This also prevents an Abelian decomposition of the SU$(2)$ fields, which only holds for homogeneous fields.  Furthermore,  dynamical gluon and photons are not treated here (in addition to the stong classical picture). Also the simple linear dependence in $z$ for the angular momentum would not fully represent the case of an HIC. We have not treated backreactions in our analysis, which could be important in early-stage dynamics; see~\cite{PhysRevD.45.4659} for a study.  The event-by-event analysis too would require an average over any given and random configuration.  In addition, the classical fields would be boost invariantly expanding~\cite{Tuchin:2013ie}, limiting the longitudinal size.  Finally, pair production can also occur for gluons~\cite{Yildiz:1979vv}, since they are self-interacting, as opposed to the Abelian photons.  Nevertheless, we will show using our simplified model, which takes inspiration from HICs, how angular momentum can arise.  But we caution that our findings, while are thought to hold more generally, have been analyzed for only the above model. A more realistic model would need to address the above limitations.

Let us spell out our main findings.   For the event-by-event averaging described above it is found that the spatial orbital angular momentum of all permutations of pair produced particles, $\boldsymbol{l}$, once averaged over is proportional to that of the background fields, namely
\begin{equation}
    \llangle\boldsymbol{l}\rrangle=\frac{|g\mathcal{B}_{\parallel}|}{8\pi^{2}}e^{-\frac{\pi m^{2}}{|g\mathcal{E}_{\parallel}|}}\coth\Bigl(\frac{\pi|\mathcal{B}_{\parallel}|}{|\mathcal{E}_{\parallel}|}\Bigr)\frac{\mathcal{E}_{\parallel}^{2}\mathcal{T}^{2}}{\mathcal{E}_{\parallel}^{2}+\mathcal{B}_{\parallel}^{2}}g^{2}\llangle\boldsymbol{l}_{\mathcal{F}}\rrangle\,.
\end{equation}
A characteristic exponential quadratic mass suppression is furthermore evident,  making visible the connection to the Schwinger effect.  While occurance of pair production is dictated by the $\mathcal{CP}$ violating parallel fields, the angular momentum dependence is entirely due to the chromoelectromagnetic momenta.  To demonstrate the above we will compute the angular momentum two ways:

\begin{enumerate}
\item The first is through an examination of Schwinger pair production occurring heuristically as a virtual breaking condensate into a particle-antiparticle pair. This approach has the virtue of physical opacity and simplicity.  We will first determine the classical equations of motion derived from the worldline for arbitrary homogeneous field.  Then we will motivate and define the heuristic model.  Finally,  we will apply event-by-event averaging.  It will be demonstrated that the physical mechanism for angular momentum inheritance stems from classical processes; however, limitations to the interpretations for the underlying physics must be drawn since the Schwinger effect is ultimately a quantum process. 
\item For second way quantum observables are addressed.  Since the Schwinger effect is inherently out-of equilibrium, we make use of the in-in formalism.  Vacuum expectation values for homogeneous fields depicted above are evaluated, and then an event-by-event averaging is applied.  In this way we confirm angular momentum transport via the Schwinger effect.
\end{enumerate}

\section{Wong's Equations: A Classical Treatment}
\label{sec:wong}

One may make use of the heuristic splitting virtual condensate picture of Schwinger pair production first by evaluating the classical equations of motions. For an SU$(2)$ two-color QCD plus QED framework, the classical equations of motion are provided by Wong's equations~\cite{Wong:1970fu,PhysRevD.17.3247,PhysRevD.15.2308}: a set of equations for a non-Abelian system resembling that of the Abelian Lorentz force. The equations describe the classical trajectories of particles with isospin (a non-Abelian charge) interacting with a  Yang-Mills field. Wong's equations directly follow from the worldline action. One may construct such an action through the one-loop effective action, whose imaginary part predicts the vacuum non-persistence of the Schwinger effect. 

To arrive at the worldline action, let us express the fermion effective action in the worldline formalism~\cite{Schubert200173}. Concretely, for partition function,
$\int\mathcal{D}\bar{\psi}\mathcal{D}\psi\exp[i\int d^{4}x\bar{\psi}(i\slashed{\mathcal{D}}-m)\psi] = \exp(i\Gamma[A,\mathcal{A}])$, one can make use of Schwinger proper time to write
\begin{align}
    \Gamma[A,\mathcal{A}]&=-\frac{i}{2}\mathrm{Tr}\ln(\slashed{\mathcal{D}}^2+m^2)
    =\frac{1}{2}\int_{0}^{\infty}\frac{dT}{T}\int d^{4}x\,\mathrm{tr}_c  \mathrm{tr}_\gamma \mathcal{K}(x,x,T)\,,
    \label{eq:eff_act}\\
    \mathcal{K}(x,y,T)&= i\langle x|e^{-i(\hat{\slashed{\mathcal{D}}}^{2}+m^{2})T}|y\rangle\,. \label{eq:kernel}
\end{align}
 $\textrm{Tr}$ is a functional trace, acting over both color and Dirac indices as well as the coordinate basis.  $\mathrm{tr}_\gamma$ is a trace over just Dirac indices.  Also, one may confirm $\slashed{\mathcal{D}}^{2}=\mathcal{D}^{2}+\frac{1}{2}(eF_{\mu\nu}+gG_{\mu\nu})\sigma^{\mu\nu}$. Equation~\eqref{eq:eff_act} may be expressed in worldline path integral form as~\cite{GSCHMIDT1993438,SCHMIDT199469,Schubert200173}
\begin{equation}
    \Gamma[A,\mathcal{A}]=\frac{i}{2}\int_{0}^{\infty}\frac{dT}{T}\oint\mathcal{{D}}x\mathcal{D}h\Phi[h] \mathrm{tr}_c \mathcal{P}_{c} \mathrm{tr}_\gamma \mathcal{P}_{\gamma}e^{-i\int_{0}^{T}d\tau[m^{2}h+\frac{1}{4h}\dot{x}^{2}+eA_\mu\dot{x}^\mu+g\mathcal{A}_\mu\dot{x}^\mu+\frac{h}{2}(eF_{\mu\nu}+gG_{\mu\nu})\sigma^{\mu\nu}]}\,.
    \label{eq:pathint0}
\end{equation}
Here the boundary conditions are understood to be $x(0)=x(T)=x'$, with path integral measure $\oint\mathcal{D}x=\int dx'\int\mathcal{D}x$. The path ordering here spans both color and Dirac indices, subscripted with $c$ and $\gamma$ respectively. Also we have included the gauge fixing functional term, $\Phi[h]$, which introduces the fluctuating variable, $h(\tau)$, into the action for a reparametrization invariant worldline expression \cite{doi:10.1142/7305}; the action is invariant under $\tau\rightarrow f(\tau)$ and $h(\tau)\rightarrow h(\tau)/f(\tau)$. We will treat the simplest gauge of $h=1$ which represents a $\delta-$functional, i.e., $\Phi[h]=\delta[h-1]$.

To arrive at Wong's equations from the action, let us express the path ordered elements as path integrals over their respective coherent states~\cite{Zhang:1990fy} or grassmann variables~\cite{doi:10.1142/7305} for color and spin degrees of freedom respectively. For the SU$(2)$ color degrees of freedom, we employ the coherent state adopted for the non-Abelian stokes theorem~\cite{PhysRevD.58.105016,PhysRevD.77.085029}. Essentially, for the coherent state, one may break up the matrix weighted propertime ordered exponential into infinitesimal discretized elements, inserting in an over-complete set of states described by a Haar measure, and sum over all possible gauge transformations. For example take $\mathcal{H}_{c}\in$ $su(2)$. Then for Haar measure, $d\mu$, of the coset SU$(2)/$U$(1)$ (spanning $S^2$), and gauge element $u\in$ SU$(2)$ one can find
\begin{equation}
    \mathrm{tr}_c \mathcal{P}_{c}e^{-ig\int_{0}^{T}d\tau\mathcal{H}_{c}}=\int\mathcal{D}\mu\exp\biggl[\frac{-ig}{2}\int_{0}^{T}d\tau \, \mathrm{tr}_c \Bigl\{\sigma_{3}\Bigl[u\mathcal{H}_{c}u^{\dagger}-\frac{i}{g}u\dot{u}^{\dagger}\Bigr]\Bigr\}\biggr]\,.
\end{equation}
By virtue of the coherent state, we may characterize our non-Abelian particles interacting with a Yang-Mills field with isospin (as introduced above, eq.~\eqref{eq:homo_fields})
\begin{equation}
    I\coloneqq \frac{1}{2}u^\dagger \sigma_3 u\,.
    \label{eq:isospin}
\end{equation}
We note that $I$ is a matrix.
Also due to the trace, we have $u(0)=u(T)$. In a similar way, one may express the spin fermionic degrees of freedom, obeying a Clifford algebra, using anti-commuting grassmann variables~\cite{doi:10.1142/7305} as
\begin{equation}
    \mathrm{tr}_\gamma \mathcal{P}_{\gamma}e^{-i\int_{0}^{T}d\tau[\frac{e}{2}F_{\mu\nu}\sigma^{\mu\nu}]}=\int\mathcal{D}\theta\exp\biggl[-i\int_{0}^{T}d\tau\Bigl(-\frac{i}{4}\theta_{\mu}\dot{\theta}^{\mu}+\frac{ie}{2}F_{\mu\nu}\theta^{\mu}\theta^{\nu}\Bigr)\biggr]\,.
\end{equation}
The above can be constructed with a complete set of states for the grassmann variables as well as a trace in the Fock space. In contrast to the color degrees of freedom for fermion statistic, we have $\{\theta_\mu,\theta_\nu\}=g_{\mu\nu}$, and we have anti-periodic boundary conditions, $\theta^{\mu}(0)=-\theta^{\mu}(T)$.
It can be seen that one may construct from the path-ordered matrix expression in eq.~\eqref{eq:pathint0} (upon infinitesimal segmentation) spanning both color and Dirac indices, the combined coherent state path integration by virtue of color and grassmann resolutions of identity. We also perform the following variable changes: $\tau \rightarrow \tau/2m$ and $T\rightarrow T/2m$. We find for the effective action, $\Gamma[A,\mathcal{A}]=(i/2)\int_{0}^{\infty}dT\,T^{-1}\oint\mathcal{D}x\mathcal{D}u\mathcal{D}\theta\,\exp(iS)$, with worldline action
\begin{equation}
    S=-\int_{0}^{T}d\tau\Bigl[ \frac{1}{2m}+\frac{m}{2}\dot{x}^{2}-\frac{i}{2} \mathrm{tr}_c (\sigma_{3}u\dot{u}^{\dagger})-\frac{i}{4}\theta_\mu\dot{\theta}^\mu+eA_\mu\dot{x}^\mu+g \mathrm{tr}_c (I \mathcal{A}_\mu)\dot{x}^\mu+\frac{i}{4m}\mathcal{F}_{\mu\nu}\theta^{\mu}\theta^{\nu}\Bigr]\,,
\end{equation}
where we have used eqs.~\eqref{eq:F_tot} and~\eqref{eq:isospin}. 

Wong's equations can be found from the worldline action. Let us first address the grassmann variable, whose equation of motion can be readily found as
\begin{equation}
    \dot{\theta}_{\mu}=\frac{1}{m}\mathcal{F}_{\mu\nu}\theta^{\nu}\,.
    \label{eq:wong_spin}
\end{equation}
To next determine the equations of motion for the coherent state representation over color, we make use of the method outlined in ref.~\cite{PhysRevD.17.3247}. We make use of the fact that $\dot{u}^{\dagger}=-u^{\dagger}\dot{u}u^{\dagger}$ and $\partial/\partial\theta_{i}u^{\dagger}=-u^{\dagger}(\partial/\partial\theta_{i}u)u^{\dagger}$. And one can eventually find for the isospin
\begin{equation}
    \dot{I}=-[ig\mathcal{A}_{\mu}\dot{x}^{\mu},I]\,,
    \label{eq:wong_isospin}
\end{equation}
where we used the fact that $\dot{I}=[I,u^{\dagger}\dot{u}]$. Conservation of isospin is guaranteed in that $2 \mathrm{tr}_c I^{2}=I^a I^a=1$ with $I=I^aT^a$. Finally, one may determine the Lorentz force non-Abelian equivalent expression in Wong's equations
\begin{equation}
    \ddot{x}_{\mu}=\frac{1}{m}\mathcal{F_{\mu\nu}}\dot{x}^{\nu}+\frac{i}{4m}\partial_\mu\mathcal{F}_{\nu\rho}\theta^\nu\theta^\rho\,.
    \label{eq:wong_lorentz}
\end{equation}
From eqs.~\eqref{eq:wong_spin} and \eqref{eq:wong_lorentz}, one can show that $\theta_\mu \dot{x}^\mu$ is  propertime independent:
\begin{equation}
\frac{d}{d\tau}(\theta_\mu \dot{x}^\mu)=\frac{1}{m}\mathcal{F}_{\mu\nu}\theta^\nu\dot{x}^\mu
+\theta^\mu\frac{1}{m}\mathcal{F}_{\mu\nu}\dot{x}^\nu
+\frac{i}{4m}\partial_\mu\mathcal{F}_{\nu\rho}\theta^\mu \theta^\nu\theta^\rho=0.
\end{equation}
Here, we employed $\theta^\mu\theta^\nu\theta^\rho\propto \epsilon^{\mu\nu\rho\sigma}$ and the Bianchi identity $\epsilon^{\mu\nu\rho\sigma}\partial_\mu F_{\nu\rho}=0$.

Let us make the connection to the Bargmann-Michel-Telegdi (BMT) equations~\cite{PhysRevLett.2.435,Skagerstam_1981}.  However, as our scope is limited to a one-loop background, let us introduce a gyromagnetic ratio phenomenologically. The discrepancy from $2$ in the magnetic moment can be entirely attributed to the coupling in the spin factor term; therefore, let us briefly digress on the case in which $\mathcal{F}_{\mu\nu}\theta^\mu\theta^\nu \rightarrow (s/2) \mathcal{F}_{\mu\nu}\theta^\mu\theta^\nu$. This would amend the spin equation of motion, eq.~\eqref{eq:wong_spin}, such that $\dot{\theta}_{\mu}=(s/2m)\mathcal{F}_{\mu\nu}\theta^{\nu}$. To construct the BMT equations, let us introduce a spin tensor~\cite{PhysRevD.16.2581} as
\begin{equation}
    S_{\mu\nu}\coloneqq-\frac{i}{2}\theta_{\mu}\theta_{\nu}\,.
\end{equation}
From which, one may find
\begin{equation}
    \dot{S}_{\mu\nu}=\frac{s}{2m}\mathcal{F}_{\mu\sigma}S_{\;\nu}^{\sigma}-\frac{s}{2m}S_{\mu}^{\;\sigma}\mathcal{F}_{\sigma\nu}\,.
    \label{eq:spin_eom}
\end{equation}
Then, let us write a Pauli-Lubanski pseudovector, or polarization tensor, as~\cite{Skagerstam_1981}
\begin{equation}
    W_{\mu}\coloneqq\frac{1}{2}\varepsilon_{\mu\nu\alpha\beta}S^{\nu\alpha}\dot{x}^{\beta}\,;
    \label{eq:pauli}
\end{equation}
note, one may also find that $\varepsilon^{\alpha\nu\mu\beta}W_{\mu}\dot{x}_{\beta}=S^{\nu\alpha}$ by choosing the vanishing constant $\theta_\mu\dot{x}^\mu=0$ and the proper time gauge $\dot{x}^2=1$. Then after some steps, one may find the BMT equation in a homogeneous background as
\begin{equation}
    \dot{W}^{\mu'}=\frac{s-2}{2m}\dot{x}^{\mu'}\mathcal{F}^{\nu\nu'}W_{\nu}\dot{x}_{\nu'}+\frac{s}{2m}\mathcal{F}^{\mu'\nu}W_{\nu}\,,
    \label{eq:BMT}
\end{equation}
in agreement with refs.~\cite{Skagerstam_1981,PhysRevD.17.3247}.

\subsection{Solutions to Wong's Equations in a SU$(2)\times$U$(1)$ Homogeneous Field}
\label{sec:wong_sol}

Furnished with Wong's equations, including the BMT equation--we however make use of the case of $s=2$, let us evaluate them exactly for SU$(2)\times$U$(1)$ homogeneous fields. To do so, it is convenient to illustrate properties of the field strength tensor, eq.~\eqref{eq:F_tot}. 

First, let us evaluate the equation of motion for isospin, eq.~\eqref{eq:wong_isospin}. The solution takes on a simple form in Abelian projected background fields. For the case of homogeneous fields in eq.~\eqref{eq:homo_fields}, one may always perform a gauge rotation to find $\mathcal{A}\propto\sigma_{3}$. The isospin, $I$, is conserved such that $2\, \mathrm{tr}_c I^{2}=1$, with $I\in$ SU$(2)/$U$(1)$, describable with a Haar measure. However, for our purposes here, the isospin will take a trivial value owing to the Abelian projection. Consider the composition of the isospin given in  eq.~\eqref{eq:isospin}, for $u\in$ SU$(2)$. A solution to Wong's equation for isospin, eq.~\eqref{eq:wong_isospin}, can be found for $u$ in the form of a Wilson loop: $u(\tau)=\mathcal{P}\exp[ig\int_{0}^{\tau}\mathcal{A}_{\mu}dx^{\mu}]u(0)$. However, since $\mathcal{A}\propto\sigma_{3}$ we can find a trivial solution of isospin that is independent of proper time as $I=(1/2)\sigma_{3}$. Therefore, we see that the isospin dynamics have been fully decoupled from the Lorentz force equation of motion, which is exactly solvable.

Let us take the case of homogeneous fields of eq.~\eqref{eq:homo_fields}. Throughout this paper where appropriate, we make use of a compact matrix notation in Lorentz indices, i.e., $\mathcal{F}\coloneqq\mathcal{F}_{\;\;\nu}^{\mu}$, $\dot{x}\coloneqq \dot{x}^\mu$ is a column vector, and we reserve the notation $\dot{x}^T\coloneqq \dot{x}_\mu$ for row vectors. Then, the Lorentz invariants, eq.~\eqref{eq:Lorentz_invariants}, follow from the field strength tensor from $\widetilde{\mathcal{F}}\mathcal{F}=I_{\tilde{F}F}\delta_{4\times4}$ and $\mathcal{F}^2-\widetilde{\mathcal{F}}^2=-I_{FF}\delta_{4\times4}$, where we have made explicit the identity matrix with $\delta_{4\times4}$. And by extension $\mathcal{F}^4+I_{FF}\mathcal{F}^2-I_{\tilde{F}F}^2\delta_{4\times 4}=0$, which actually follows from Cayley-Hamilton's theoreom since $\det(\mathcal{F}-\lambda \delta_{4\times 4})=\lambda^{4}+\lambda^{2}I_{FF}-I_{\tilde{F}F}^{2}=0$~\cite{Fradkin_1978}. Then the eigenvalues of the field strength tensor can found as $\pm\lambda_{E}$ and $\pm i\lambda_{B}$ with respective magnitudes
\begin{equation}
    \lambda_{E}\coloneqq\frac{1}{\sqrt{2}}\sqrt{\sqrt{I_{FF}^{2}+4I_{\tilde{F}F}^{2}}-I_{FF}}\,,\quad \lambda_{B}\coloneqq\frac{1}{\sqrt{2}}\sqrt{\sqrt{I_{FF}^{2}+4I_{\tilde{F}F}^{2}}+I_{FF}}\,,
    \label{eq:F_eigenvalues}
\end{equation}
which are respectively projections of the electric and magnetic field strengths. The magnitudes of the eigenvalues satisfy the following identities in terms of the Lorentz invariants: $\lambda_{E}\lambda_{B}=|I_{\tilde{F}F}|$, and $\lambda_{B}^{2}-\lambda_{E}^{2}=I_{FF}$.

It is also convenient to perform the full eigendecomposition of the field strength tensor; we use the approach used in ref.~\cite{Fradkin_1978}. Most relevant are the projection operators of the squared eigenvalues, given in eq.~\eqref{eq:F_eigenvalues}, which can be readily verified as
\begin{equation}
   P_{E}\coloneqq\frac{\lambda_{B}^{2}+\mathcal{F}^{2}}{\lambda_{E}^{2}+\lambda_{B}^{2}}\,,\quad P_{B}\coloneqq\frac{\lambda_{E}^{2}-\mathcal{F}^{2}}{\lambda_{E}^{2}+\lambda_{B}^{2}}\,,
   \label{eq:proj_op}
\end{equation}
which satisfy $\mathcal{F}^{2}P_{E}=\lambda_{E}^{2}P_{E}$, and $\mathcal{F}^{2}P_{B}=-\lambda_{B}^{2}P_{B}$. The projection operators are also idempotent, complete, and orthogonal: $P_{E}^{2}=P_{E}$, $P_{B}^{2}=P_{B}$, $P_{E}+P_{B}=1$, and  $P_{E}P_{B}=0$. And later, it will be necessary to construct the eigenvectors of the tensor (we use conventions in ref.~\cite{etde_6972673}), and they can be verified as
\begin{equation}
    n_{E}^{\pm}=(\mathcal{F}\pm\lambda_{E})P_{E}\xi_{E}^{\pm}\,,\quad n_{B}^{\pm}=(\mathcal{F}\pm i\lambda_{B})P_{B}\xi_{B}^{\pm}\,,
    \label{eq:F_eigenvectors}
\end{equation}
which satisfy $\mathcal{F}n_{E}^{\pm}=\pm\lambda_{E}n_{E}^{\pm}$, and $\mathcal{F}n_{B}^{\pm}=\pm i\lambda_{B}n_{B}^{\pm}$. They are also orthogonal since $n_{E}^{\pm\,T}n_{E}^{\pm}=n_{B}^{\pm\,T}n_{B}^{\pm}=0$, and $n_{E}^{T}n_{B}=n_{B}^{T}n_{E}=0$. $\xi_E^\pm$ and $\xi_B^\pm$ may be selected for normalization such that $n_{E}^{-\,T}n_{E}^{+}=n_{E}^{+\,T}n_{E}^{-}=2$, and  $n_{B}^{-\,T}n_{B}^{+}=n_{B}^{+\,T}n_{B}^{-}=-2$; however, their specific form is unimportant for most of our purposes, in the instances where important we will use parallel electric and magnetic fields that have a physically transparent eigenvector construction. Rather than the eigenvectors given in eq.~\eqref{eq:F_eigenvectors}, it is convenient to use the following combinations:
\begin{align}
    n_{\lambda_{E}}^{+} & =\frac{1}{2}(n_{E}^{-}+n_{E}^{+})\,,\quad n_{\lambda_{E}}^{-}=\frac{1}{2}(n_{E}^{-}-n_{E}^{+})\,,\\
    n_{\lambda_{B}}^{+} & =\frac{1}{2}(n_{B}^{-}+n_{B}^{+})\,,\quad n_{\lambda_{B}}^{-}=\frac{1}{2}(n_{B}^{-}-n_{B}^{+})\,;
    \label{eq:F_comb_eigenvectors}
\end{align}
the above combinations satisfy
\begin{equation}
    \mathcal{F}n_{\lambda_{E}}^{\pm} =-\lambda_{E}n_{\lambda_{E}}^{\mp}\,, \quad \mathcal{F}n_{\lambda_{B}}^{\pm}=-i\lambda_{B}n_{\lambda_{B}}^{\mp}\,. \label{eq:Fn}
\end{equation}
They are normalized such that $n_{\lambda_{E}}^{+\,T}n_{\lambda_{E}}^{+} =1$, $ n_{\lambda_{E}}^{-\,T}n_{\lambda_{E}}^{-}=-1$, $n_{\lambda_{B}}^{+\,T}n_{\lambda_{B}}^{+}=-1$, and $n_{\lambda_{B}}^{-\,T}n_{\lambda_{B}}^{-}=1$. $n_{\lambda_E}^+$ is time-like and thus its contraction with a vector represents the time component; for example for the case of a pure electric field in the $\hat{x}_3$ direction, $n_{\lambda_E}^{+\,T}x$ would be $x_0$ whereas $n_{\lambda_E}^{-\,T}x$ would be $\mathrm{sgn}(g\mathcal{E}) x_3$. Let us also mention that the fully antisymmetric field strength tensor has a similar eigendecomposition: for
$\widetilde{\mathcal{F}}n_{E}^{\pm}=\pm(I_{\tilde{F}F}/\lambda_{E})n_{E}^{\pm}$, and $\widetilde{\mathcal{F}}n_{B}^{\pm}=\mp i(I_{\tilde{F}F}/\lambda_{B})n_{B}^{\pm}$, one can also find
\begin{equation}
    \widetilde{\mathcal{F}}n_{\lambda_{E}}^{\pm} =-\frac{I_{\tilde{F}F}}{\lambda_{E}}n_{\lambda_{E}}^{\mp}\,,\quad \widetilde{\mathcal{F}}n_{\lambda_{B}}^{\pm}=i\frac{I_{\tilde{F}F}}{\lambda_{B}}n_{\lambda_{B}}^{\mp}\,. \label{eq:tildeFn}
\end{equation}
The combined eigenvectors in eq.~\eqref{eq:F_comb_eigenvectors} allow one to perform the spectral decomposition of the field strength tensors as
\begin{align}
    \mathcal{F}&=\lambda_{E}[n_{\lambda_{E}}^{+}n_{\lambda_{E}}^{-\,T}-n_{\lambda_{E}}^{-}n_{\lambda_{E}}^{+\,T}]-i\lambda_{B}[n_{\lambda_{B}}^{+}n_{\lambda_{B}}^{-\,T}-n_{\lambda_{B}}^{-}n_{\lambda_{B}}^{+\,T}]\,,\\ \widetilde{\mathcal{F}}&=\frac{I_{\tilde{F}F}}{\lambda_{E}}[n_{\lambda_{E}}^{+}n_{\lambda_{E}}^{-\,T}-n_{\lambda_{E}}^{-}n_{\lambda_{E}}^{+\,T}]+i\frac{I_{\tilde{F}F}}{\lambda_{B}}[n_{\lambda_{B}}^{+}n_{\lambda_{B}}^{-\,T}-n_{\lambda_{B}}^{-}n_{\lambda_{B}}^{+\,T}]\,.
    \label{eq:F_spectral}
\end{align}
Finally, it is occasionally convenient to represent the projection operators, eq.~\eqref{eq:proj_op}, using the combined eigenvectors as
\begin{equation}
     P_{E}= n_{\lambda_{E}}^{+}n_{\lambda_{E}}^{+\,T}-n_{\lambda_{E}}^{-}n_{\lambda_{E}}^{-\,T}\,,\quad P_{B} = n_{\lambda_{B}}^{-}n_{\lambda_{B}}^{-\,T}-n_{\lambda_{B}}^{+}n_{\lambda_{B}}^{+\,T}\,.
     \label{eq:proj_operators}
\end{equation}

Using the above eigendecomposition, particularly using the projection operators, one may readily solve the Lorentz force equation, eq.~\eqref{eq:wong_lorentz}, in homogeneous fields~\cite{Fradkin_1978}; see also ref.~\cite{PhysRevD.103.036004}. To do so, let us define $q$ such that $\dot{x}=:(d/d\tau+m^{-1}\mathcal{F})q$, with $q_{E}\coloneqq P_{E}q$, and $q_{B}\coloneqq P_{B}q$. Then usage of $q_E$ and $q_B$ enables the Lorentz force equation to be decoupled as
\begin{equation}
    \Bigl(\frac{d^{2}}{d\tau^{2}}-\frac{\lambda_{E}^{2}}{m^2}\Bigr)q_{E}=0\,,\quad\Bigl(\frac{d^{2}}{d\tau^{2}}+\frac{\lambda_{B}^{2}}{m^2}\Bigr)q_{B}=0\,.
\end{equation}
Using the fact that $P_Bq_E=0$ and $P_Eq_B=0$, one can evaluate the two above equations separately, then combine to find
\begin{equation}
    \dot{x}(\tau)=\biggl\{\Bigl[\cosh\Bigl(\frac{\lambda_{E}\tau}{m}\Bigr)+\lambda_{E}^{-1}\mathcal{F}\sinh\Bigl(\frac{\lambda_{E}\tau}{m}\Bigr)\Bigr]P_E +\Bigl[\cos\Bigl(\frac{\lambda_{B}\tau}{m}\Bigr)+\lambda_{B}^{-1}\mathcal{F}\sin\Bigl(\frac{\lambda_{B}\tau}{m}\Bigr)\Bigr]P_B \biggr\}\dot{x}(0)\,.
    \label{eq:velocity_sol}
\end{equation}
Note, the expression in the curly brackets agrees with the exponential of the field strength tensor such that $\dot{x}(\tau)=\exp(\mathcal{F}\tau/m)\dot{x}(0)$. We have also arbitrarily chosen an initial proper time of $\tau_0 = 0$. Last, the coordinate solution can be verified as
\begin{equation}
    x(\tau)=\frac{m}{I_{\tilde{F}F}}\widetilde{\mathcal{F}}[\dot{x}(\tau)-\dot{x}(0)]+x(0)\,.
    \label{eq:coordinate_sol}
\end{equation}

Not only the Lorentz force equation, but also the BMT equation, can one find an exact solution in homogeneous fields. A general solution to the spin tensor equation given in eq.~\eqref{eq:spin_eom} can be found as $S(\tau)=\exp[(s/2m)\mathcal{F}\tau]S(0)\exp [-(s/2m)\mathcal{F}\tau]$. Then let us introduce the fully anti-symmetric spin tensor as
\begin{equation}
    \widetilde{S}=I_{\tilde{S}S}S^{-1}\,,\quad I_{\tilde{S}S}=-\frac{1}{4}\widetilde{S}_{\mu\nu}S^{\mu\nu}\,,
\end{equation}
from which one may directly write down the general solution to the BMT equation, eq.~\eqref{eq:BMT}, as
\begin{equation}
    W(\tau)=\widetilde{S}(\tau)\dot{x}(\tau)\,,\quad \widetilde{S}(\tau) =e^{\frac{s}{2m}\mathcal{F}\tau}\widetilde{S}(0)e^{-\frac{s}{2m}\mathcal{F}\tau}\,,
\end{equation}
with $\dot{x}(\tau)$ given by eq.~\eqref{eq:velocity_sol}. Let us however confine our attention to the case of a gyromagnetic ratio of $s=2$, which will simplify matters: $W(\tau)=\exp(\mathcal{F}\tau/m)W(0)$ for some initial polarization, $W(0)$. Note that the selection of the gyromagnetic ratio entails the constraint $\dot{x}_{\mu}\theta^{\mu}=0$. The solution for the Pauli-Lubanski pseudovector can now be found similar as was found for as eq.~\eqref{eq:velocity_sol}:
\begin{equation}
    W(\tau)=\biggl\{\Bigl[\cosh\Bigl(\frac{\lambda_{E}\tau}{m}\Bigr)+\lambda_{E}^{-1}\mathcal{F}\sinh\Bigl(\frac{\lambda_{E}\tau}{m}\Bigr)\bigr]P_E +\Bigl[\cos\Bigl(\frac{\lambda_{B}\tau}{m}\Bigr)+\lambda_{B}^{-1}\mathcal{F}\sin\Bigl(\frac{\lambda_{B}\tau}{m}\Bigr)\Bigr]P_B \biggr\} W(0)\,.
    \label{eq:solution_pauli}
\end{equation}

\subsection{Angular Momentum from a Splitting Virtual Condensate}
\label{sec:heuristic}

Before we address the full one-loop quantum calculation of angular momentum via the Schwinger effect, let us first ensure its validity in the classical picture of a virtual splitting condensate. While the picture is indeed only heuristic, values derived from the picture agree well with those of expectation values of one-loop background calculations in homogeneous fields~\cite{PhysRevLett.104.212001}. Let us describe the physical picture: Envision a pair of particles produced from the vacuum, which we characterize as a virtual particle-antiparticle condensate, in some point in space time. 
In the heuristic model we confine our attention to the $n=1$ instanton contribution to the non-persistence probability~\cite{PhysRevD.78.036008}.  This is because the heuistic picture treats only a single pair.  Even so,  the treatment is not limited to the case of just weak fields. Indeed we will find in the following sections when computing in-in oberservables to all $n$ orders that similar expectation values can be had for fields with arbitrary strengths.  For our SU$(2)\times$U$(1)$ homogeneous fields one can find for the imaginary part of the effective action, eq.~\eqref{eq:pathint0}, the following non-persistence probability per unit volume-time of a single particle-antiparticle pair with opposite charge and color as~\cite{PhysRevLett.104.212001}
\begin{equation}
    \mathcal{W}=\sum_{\pm g}\frac{\lambda_E \lambda_B \mathcal{V}\mathcal{T}}{4\pi^2}\coth\Bigl(\frac{\pi \lambda_B}{\lambda_E}\Bigr)\exp\Bigl(-\frac{\pi m^{2}}{\lambda_{E}}\Bigr)\,.
    \label{eq:nonpersistence}
\end{equation}
Here $\mathcal{V}$ and $\mathcal{T}$ are the system volume and real time, these formal divergent factors arise in canonical momentum integrals in the effective action, and are a result of infinitely spanning homogeneous fields~\cite{Nikishov:1969tt}. A concrete form for the kernel, eq.~\eqref{eq:kernel_soln}, will be explicitly provided later from which the above simply follows from the $n=1$ pole on the imaginary proper time axis. Upon creation it is assumed the particle-antiparticle pair evolve for a long real time, $\sim\mathcal{T}$, classically according to Wong's equations (including the BMT equation), eqs.~\eqref{eq:wong_isospin},~\eqref{eq:wong_lorentz}, and~\eqref{eq:BMT}. Last, two types of pairs of particles may be produced: both the pair $(e,g)$ and $(-e,-g)$ as well as the pair $(e,-g)$ and $(-e,g)$. However, while the sum of both probabilities of such pairs is encoded into the non-persistence, eq.~\eqref{eq:nonpersistence}, their respective probabilities cannot be easily decoupled. Nevertheless, let us consider the HIC setup in section~\ref{sec:hic}; there one can see that $\lambda_E,\lambda_B|_{g\rightarrow+g}\approx\lambda_E,\lambda_B|_{g\rightarrow-g}$ for fixed U$(1)$ coupling. Furthermore, the non-persistence is positive definite. Therefore let us treat the total non-persistence as
\begin{equation}
    \mathcal{W}\approx\frac{\lambda_E \lambda_B \mathcal{V}\mathcal{T}}{2\pi^2}\coth\Bigl(\frac{\pi \lambda_B}{\lambda_E}\Bigr)\exp\Bigl(-\frac{\pi m^{2}}{\lambda_{E}}\Bigr)\,,
    \label{eq:nonpersistence2}
\end{equation}
which holds for either $\pm g$. With the above non-persistence, either pair will in fact give an identical rate of occurrence for our fields considered. Last, there is a factor of two in the above (in contrast to the strictly Abelian case) since the rate of occurrence of both particle pairs is taken into account.

For the heuristic picture, let us restrict our attention to the initial condition of rest: $\dot{x}_{\mu}(0)=(1,0,0,0)$ for either particle or antiparticle. Particles being produced with finite momentum are exponentially suppressed, and thus this serves as a good approximation. Then it is convenient for the following calculations to write out the solution to the Lorentz force equation. The velocity, eq.~\eqref{eq:velocity_sol}, and coordinates, eq.~\eqref{eq:coordinate_sol}, become
\begin{align}
    \dot{x}^{0}(\tau)&=\frac{1}{\lambda_{E}^{2}+\lambda_{B}^{2}}\biggl\{(\lambda_{B}^{2}+g^{2}\mathcal{E}^{2})\cosh\Bigl(\frac{\lambda_{E}\tau}{m}\Bigr)+(\lambda_{E}^{2}-g^{2}\mathcal{E}^{2})\cos\Bigl(\frac{\lambda_{E}\tau}{m}\Bigr)\biggr\}\,,\\
    \dot{\boldsymbol{x}}(\tau)&=\frac{1}{\lambda_{E}^{2}+\lambda_{B}^{2}}\biggl\{\Bigl[\cosh\Bigl(\frac{\lambda_{E}\tau}{m}\Bigr)-\cos\Bigl(\frac{\lambda_{B}\tau}{m}\Bigr)\Bigr]g\boldsymbol{\mathcal{E}}\times(g\boldsymbol{\mathcal{B}}+e\boldsymbol{B})+\Bigl[\lambda_{E}\sinh\Bigl(\frac{\lambda_{E}\tau}{m}\Bigr)\nonumber \\
    &+\lambda_{B}\sin\Bigl(\frac{\lambda_{B}\tau}{m}\Bigr)\Bigr]g\boldsymbol{\mathcal{E}}+I_{\tilde{F}F}\Bigl[\lambda_{E}^{-1}\sinh\Bigl(\frac{\lambda_{E}\tau}{m}\Bigr)-\lambda_{B}^{-1}\sin\Bigl(\frac{\lambda_{B}\tau}{m}\Bigr)](g\boldsymbol{\mathcal{B}}+e\boldsymbol{B})\biggr\}\,,\\
    x^{0}(\tau)&=\frac{m}{\lambda_{E}^{2}+\lambda_{B}^{2}}\biggl\{\frac{\lambda_{B}^{2}+g^{2}\mathcal{E}^{2}}{\lambda_{E}}\sinh\Bigl(\frac{\lambda_{E}\tau}{m}\Bigr)+\frac{\lambda_{E}^{2}-g^{2}\mathcal{E}^{2}}{\lambda_B}\sin\Bigl(\frac{\lambda_{B}\tau}{m}\Bigr)\biggr\}+x^{0}(0)\,,\\
    \boldsymbol{x}(\tau)&=\frac{m}{\lambda_{E}^{2}+\lambda_{B}^{2}}\biggl\{\Bigl[\lambda_{E}^{-1}\sinh\Bigl(\frac{\lambda_{E}\tau}{m}\Bigr)-\lambda_{B}^{-1}\sin\Bigl(\frac{\lambda_{B}\tau}{m}\Bigr)\Bigr]g\boldsymbol{\mathcal{E}}\times(g\boldsymbol{\mathcal{B}}+e\boldsymbol{B})+\Bigl[\frac{\lambda_{B}^{2}}{I_{\tilde{F}F}}\Bigl(\cosh\Bigl(\frac{\lambda_{E}\tau}{m}\Bigr)-1\Bigr)\nonumber\\
    &+\frac{\lambda_{E}^{2}}{I_{\tilde{F}F}}\Bigl(\cos\Bigl(\frac{\lambda_{B}\tau}{m}\Bigr)-1\Bigr)\Bigr](g\boldsymbol{\mathcal{B}}+e\boldsymbol{B})+\Bigl[\cosh\Bigl(\frac{\lambda_{E}\tau}{m}\Bigr)-\cos\Bigl(\frac{\lambda_{B}\tau}{m}\Bigr)\Bigr]g\boldsymbol{\mathcal{E}}\biggr\}+\boldsymbol{x}(0)\,.
    \label{eq:coordinate}
\end{align}

To show the heuristic picture and to ensure its validity, we demonstrate a simple calculation of the electromagnetic vector current in a sole electric field in the $\hat{\boldsymbol{x}}_{E}$ direction. In the heuristic picture one may show that for a particle-antiparticle pair with $(\pm e,\pm g)$
\begin{align}
    \boldsymbol{J}_{(\pm e,\pm g)}&= \frac{e}{2}\mathcal{W}\int d^3y\int d\tau\,\Bigl\{\bigl[\delta^{4}(y-x(\tau))\dot{\boldsymbol{x}}(\tau)\big|_{e,g\rightarrow e,g}-\bigl[\delta^{4}(y-x(\tau))\dot{\boldsymbol{x}}(\tau)\big|_{e,g\rightarrow-e,-g}\Bigr\} \notag\\
    &=\frac{e}{2}\mathcal{W}\Bigl\{\Bigl[\frac{\dot{\boldsymbol{x}}(y_0)}{|\dot{x}^{0}(y_0)|}\Big|_{e,g\rightarrow e,g}-\Bigl[\frac{\dot{\boldsymbol{x}}(y_0)}{|\dot{x}^{0}(y_0)|}\Big|_{e,g\rightarrow-e,-g}\Bigr\}\,;
\end{align}
likewise one may write a similar current for the pair $(\pm e,\mp g)$. Here $y_0\sim \mathcal{T}/2 \gg 0$. Note that we have a factor of $1/2$ in eq.~\eqref{eq:nonpersistence2} to account for a single species of pairs. Let us also point out that for the Lorentz invariants one finds
\begin{equation}
    I_{\tilde{F}F},\,I_{FF},\,\lambda_{E},\,\lambda_{B}\big|_{e,g\rightarrow e,g} =I_{\tilde{F}F},\,I_{FF},\,\lambda_{E},\,\lambda_{B}\big|_{e,g\rightarrow-e,-g}\,.
\end{equation}
To simplify this cursory discussion, let us examine the case with only an electric field in the direction $\hat{\boldsymbol{x}}_{E}$, whose eigenvalues are simply $\lim_{B\rightarrow0}\lambda_{E}=|g\mathcal{E}|$, and $\lim_{B\rightarrow0}\lambda_{B}=0$. Then for large $y^0=(m/g\mathcal{E})\sinh(g\mathcal{E}\tau/m)$ one will find as anticipated that the generated electromagnetic current of the $(\pm e,\pm g)$ pair will saturate to 
$2e(1/2)\mathcal{W}$~\cite{PhysRevLett.104.212001}:
\begin{equation}
    \boldsymbol{J}_{(\pm e,\pm g)}=\mathrm{sgn}(g)\frac{e}{2}\frac{\lambda_E^2 \mathcal{V}\mathcal{T}}{\pi^3}\exp\Bigl(-\frac{\pi m^{2}}{\lambda_{E}}\Bigr)\hat{\boldsymbol{x}}_{E}\,.
    \label{eq:current_heuristic}
\end{equation}
Importantly, there is an overall dependence on the sign of $g$ in the vector current. Therefore for the contribution from the other pair, we find that $\boldsymbol{J}_{(\pm e,\pm g)}=-\boldsymbol{J}_{(\pm e,\mp g)}$, and hence the total current, $\boldsymbol{J}_{(\pm e,\pm g)}+\boldsymbol{J}_{(\pm e,\mp g)}$, is vanishing. That there can be no electromagnetic current from the Schwinger effect in SU$(2)\times$U$(1)$ due to the cancellation from both color contributions has been studied in ref.~\cite{PhysRevD.92.125012}.   Using the heuristic model,  it appears as though transport of current (and other observables) occurs from classical processes instead of the Schwinger effect, beginning at time $\mathcal{T}/2=0$.  However, $\mathcal{T}$ is also the system duration,  and depicts the time the electric field is turned on, and hence $\mathcal{T}/2=0$ depicts a scenario with no electric field.  While the heuristic model is powerful and intuitive, it is still a classical picture, while the Schwinger effect is a quantum phenomenon, and there are limitations we can draw concerning the underlying physics.  We can demonstrate transport of the current as well as angular momentum using the fully quantum in-in contruction in the following sections. Let us next determine the orbital and spin angular momentum using the heuristic picture.

Let us first address the spin in the heuristic picture governed by the BMT equation, eq.~\eqref{eq:BMT}. The initial condition of rest for the particle-antiparticle pair also imposes a restriction on the initial Pauli-Lubanski pseudovector, i.e., from eq.~\eqref{eq:pauli} one finds $W_{\mu}(0)=\widetilde{S}_{\mu0}(0)$. For the totally antisymmetric spin tensor, this corresponds to the magnetic-like part of the spin tensor. Finally note that in the worldline action the coupled spin tensor to field strength tensor term,  $(1/2)\mathcal{F}_{\mu\nu}S^{\mu\nu}$,  indicates that the energy is minimized for spin alignment with the magnetic field. Therefore, we take for our initial spin state as
$\boldsymbol{W}(0)=(e\boldsymbol{B}+g\boldsymbol{\mathcal{B}})/|e\boldsymbol{B}+g\boldsymbol{\mathcal{B}}|$,
and hence the polarization vector, eq.~\eqref{eq:solution_pauli}, becomes
\begin{align}
    W^{0}(\tau)&=\frac{I_{\tilde{F}F}}{\lambda_{E}^{2}+\lambda_{B}^{2}}\biggl\{\Bigl[\lambda_{E}\sinh\Bigl(\frac{\lambda_{E}\tau}{m}\Bigr)+\lambda_{B}\sin\Bigl(\frac{\lambda_{B}\tau}{m}\Bigr)\Bigr]\frac{1}{|e\boldsymbol{B}+g\boldsymbol{\mathcal{B}}|}\nonumber \\
    &+\Bigl[\lambda_{E}^{-1}\sinh\Bigl(\frac{\lambda_{E}\tau}{m}\Bigr)-\lambda_{B}^{-1}\sin\Bigl(\frac{\lambda_{B}\tau}{m}\Bigr)\Bigr]|e\boldsymbol{B}+g\boldsymbol{\mathcal{B}}|\biggr\}\,,
    \\\boldsymbol{W}(\tau)&=\frac{1}{\lambda_{E}^{2}+\lambda_{B}^{2}}\frac{1}{|e\boldsymbol{B}+g\boldsymbol{\mathcal{B}}|}\biggr\{\Bigl[\lambda_{B}^{2}\cosh\Bigl(\frac{\lambda_{E}\tau}{m}\Bigr)+\lambda_{E}^{2}\cos\Bigl(\frac{\lambda_{B}\tau}{m}\Bigr)\Bigr](e\boldsymbol{B}+g\boldsymbol{\mathcal{B}})+I_{\tilde{F}F} \Bigl\{\Bigl[\cosh\Bigl(\frac{\lambda_{E}\tau}{m}\Bigr)\nonumber \\
    &-\cos\Bigl(\frac{\lambda_{B}\tau}{m}\Bigr)\Bigr]g\boldsymbol{\mathcal{E}}+ \Bigl[\lambda_{E}^{-1}\sinh\Bigl(\frac{\lambda_{E}\tau}{m}\Bigr)-\lambda_{B}^{-1}\sin\Bigl(\frac{\lambda_{B}\tau}{m}\Bigr)\Bigr]g\boldsymbol{\mathcal{E}}\times(e\boldsymbol{B}+g\boldsymbol{\mathcal{B}})\Bigr\}\biggr\}\,.
\end{align}
Here, the produced particle-antiparticle pair total pseudovector follows similar to the current for the $(\pm e,\pm g)$ pair as
\begin{equation}
    P^{\mu}_{(\pm e,\pm g)}=\frac{\mathcal{W}}{2}\Bigl\{\Bigl[\frac{W^{\mu}(y_{0})}{|\dot{x}^{0}(y_{0})|}\Big|_{e,g\rightarrow e,g}+\Bigl[\frac{W^{\mu}(y_{0})}{|\dot{x}^{0}(y_{0})|}\Big|_{e,g\rightarrow-e,-g}\Bigr\}\,.
    \label{eq:P}
\end{equation}
Specializing to the fields as described in section~\ref{sec:hic} one can find for the temporal and spatial components as
\begin{align}
    P_{(\pm e,\pm g)}^{0}&\approx g^{2}\boldsymbol{\mathcal{E}}_{\parallel}\cdot\boldsymbol{\mathcal{B}}_{\parallel}\frac{|\mathcal{E}_{\parallel}\mathcal{B}_{\parallel}|\mathcal{T}\mathcal{V}}{2\pi^{2}(\mathcal{B}_{\parallel}^{2}+\mathcal{E}_{\parallel}^{2})}\coth\biggl(\frac{|\mathcal{B}_{\parallel}|\pi}{|\mathcal{E}_{\parallel}|}\biggr)e^{-\frac{m^{2}\pi}{|g\mathcal{E}_{\parallel}|}}\biggl\{\frac{|\mathcal{E}_{\parallel}|}{|\boldsymbol{\mathcal{B}}|}+\frac{|\boldsymbol{\mathcal{B}}|}{|\mathcal{E}_{\parallel}|}\biggr\}=P_{(\pm e,\mp g)}^{0}\,,\\
    \boldsymbol{P}_{(\pm e,\pm g)}&\approx g^{2}\boldsymbol{\mathcal{E}}_{\parallel}\cdot\boldsymbol{\mathcal{B}}_{\parallel}\frac{\mathcal{T}\mathcal{V}}{2\pi^{2}}\coth\biggl(\frac{|\mathcal{B}_{\parallel}|\pi}{|\mathcal{E}_{\parallel}|}\biggr)e^{-\frac{m^{2}\pi}{|g\mathcal{E}_{\parallel}|}}\frac{|\mathcal{B}_{\parallel}|}{|\boldsymbol{\mathcal{B}}|(\mathcal{B}_{\parallel}^{2}+\mathcal{E}_{\parallel}^{2})}\notag \\
    &\times [ \boldsymbol{\mathcal{E}}_z \times \boldsymbol{\mathcal{B}}_z + \boldsymbol{\mathcal{E}}_z \times\boldsymbol{\mathcal{B}}_{\parallel}+\boldsymbol{\mathcal{E}}_{\parallel}\times \boldsymbol{\mathcal{B}}_z ]=\boldsymbol{P}_{(\pm e,\mp g)}\,.
\end{align}
However, after averaging over $\pm \mathcal{E}_\parallel$ and $\pm \mathcal{B}_\parallel$ in eq.~\eqref{eq:hic_event}, we can determine that the relevant observables, i.e., $\llangle P^0 \rrangle$ and $\llangle \boldsymbol{z} \times \boldsymbol{P}\rrangle$, will vanish. Thus one may rule out spin as a contributor to the total angular momentum here. This is however to be expected. Since the pseudovector can be likened to a chiral vector current that is proportional to $I_{\tilde{F}F}$, which will vanish according to an averaging over all events as assumed in eq.~\eqref{eq:hic_event}.

To further illustrate this point let us look at a scenario different than that given in section~\ref{sec:hic} with parallel electric and magnetic fields such that $\boldsymbol{l}_{\mathcal{F}}=0$--see eq.~\eqref{eq:F_angular_density}--and $I_{\tilde{F}F}\neq0$. Let us also treat a strong magnetic field such that only the lowest Landau level would be present in effect polarizing the particles' spins. Then using eq.~\eqref{eq:P}, we can find that
\begin{equation}
    P^{0}_{\pm e, \pm g}=\frac{\mathcal{W}I_{\tilde{F}F}}{\lambda_{B}^{2}+g^{2}\mathcal{E}^{2}}\Bigl\{\frac{\lambda_{E}}{|e\boldsymbol{B}+g\boldsymbol{\mathcal{B}}|}+\frac{|e\boldsymbol{B}+g\boldsymbol{\mathcal{B}}|}{\lambda_{E}}\Bigr\}\approx I_{\tilde{F}F}\frac{\mathcal{T}\mathcal{V}}{2 \pi^{2}}e^{-\frac{\pi m^{2}}{\lambda_{E}}}\,.
\end{equation}
Furthermore since $P^{0}_{\pm e, \pm g}=P^{0}_{\pm e, \mp g}$ (in contrast to the vector current found above) one can find that the non-vanishing time component total pseudovector in parallel fields becomes
\begin{equation}
    P^{0}=P^{0}_{\pm e, \pm g}+P^{0}_{\pm e, \mp g} = I_{\tilde{F}F}\frac{\mathcal{T}\mathcal{V}}{\pi^{2}}e^{-\frac{\pi m^{2}}{\lambda_{E}}}\,.
    \label{eq:p_0_total}
\end{equation}
If one were to interpret the total time of the system, $\mathcal{T}$, as a differentiable real-time, and furthermore liken the time-like pseudovector to a chiral density, then the above would resemble the chiral anomaly in the massless limit. As our background field's total angular momentum goes as $\boldsymbol{\mathcal{E}}\times \boldsymbol{\mathcal{B}}$, c.f., eq.~\eqref{eq:F_total_ang_mom}, one should expect to see pairs with such an angular momentum dependence, and we can show this is the case with the orbital part. 

For a single point-like particle, we require a Lorentz covariant definition of the particle momentum, which stems from the energy-momentum tensor density~\cite{barut1980electrodynamics},  $m\int_{-\infty}^{\infty}d\tau\,\delta(y-x(\tau))\dot{x}^{\mu}(\tau)\dot{x}^{\nu}(\tau)$. And we have for the combined point-like particle energy-momentum tensor for the $(\pm e, \pm g)$ pair as
\begin{equation}
    \mathcal{T}^{\mu\nu}_{(\pm e, \pm g)}=\frac{m\mathcal{W}}{2}\Bigl\{\Bigl[\frac{\dot{x}^{\mu}(y_{0})\dot{x}^{\nu}(y_{0})}{|\dot{x}^{0}(y_{0})|}\Big|_{e,g\rightarrow e,g}+\Bigl[\frac{\dot{x}^{\mu}(y_{0})\dot{x}^{\nu}(y_{0})}{|\dot{x}^{0}(y_{0})|}\Bigr|_{e,g\rightarrow -e,-g}\Bigr\}\,,
    \label{eq:heur_tensor_def}
\end{equation}
whose $\mathcal{T}^{0i}$ components are the spatial momentum. However, here in contrast to the vector current and polarization, we see the covariant spatial momentum grows (apart from the pair production rate, $\mathcal{W}$, factor) with time, and hence is quadratic with time with the $\mathcal{W}$ factor. The time growth and hence system real-time can be determined as
\begin{equation}
    x^0(\tau\gg0)=\frac{m}{\lambda_{E}}\frac{\lambda_{B}^{2}+g^{2}\mathcal{E}^{2}}{\lambda_{E}^{2}+\lambda_{B}^{2}}\sinh\Bigl(\frac{\lambda_{E}\tau}{m}\Bigr) \rightarrow \frac{\mathcal{T}}{2}\,;
\end{equation}
Here the right arrow follows after the delta function takes $x^0\rightarrow y^0$. Using the conditions dictated in section~\ref{sec:hic},  one may find for the combined momentum as
\begin{equation}
    \mathcal{T}^{0i}_{(\pm e, \pm g )}=\frac{\mathcal{T}^{2}\mathcal{V}}{4\pi^{2}}\frac{\mathcal{E}_{\parallel}^{2}|g\mathcal{B}_{\parallel}|}{\mathcal{B}_{\parallel}^{2}+\mathcal{E}_{\parallel}^{2}}\coth\biggl(\frac{|\mathcal{B}_{\parallel}|\pi}{|\mathcal{E}_{\parallel}|}\biggr)e^{-\frac{m^{2}\pi}{|g\mathcal{E}_{\parallel}|}}g^{2}[ \boldsymbol{\mathcal{E}}_z \times \boldsymbol{\mathcal{B}}_z + \boldsymbol{\mathcal{E}}_z \times\boldsymbol{\mathcal{B}}_{\parallel}+\boldsymbol{\mathcal{E}}_{\parallel}\times \boldsymbol{\mathcal{B}}_z ]\,.
    \label{eq:heuristic_momentum}
\end{equation}
Thus, we see as expected a net momentum associated with pair production. Finally, we note that $\mathcal{T}^{0i}_{(\pm e, \pm g )}=\mathcal{T}^{0i}_{(\pm e, \mp g )}$.

The total angular momentum density is then
\begin{equation}
    L^{\mu\nu\sigma}=x^{\nu}\mathcal{T}^{\mu\sigma}-x^{\sigma}\mathcal{T}^{\mu\nu}\,.
    \label{eq:angmomclassfinal}
\end{equation}
Here $x^{\nu}$ and $x^\sigma$ are understood to act on both the $(\pm e, \pm g)$ and $(\pm e, \mp g)$ pairs and are given by eq.~\eqref{eq:coordinate}. We take the initial criteria as $\boldsymbol{x}(0)|_{e,g\rightarrow e,g}=\boldsymbol{x}(0)|_{e,g\rightarrow-e,-g}=\boldsymbol{z}$ as illustrated in section~\ref{sec:hic}, and hence both particles originating at a similar point.
We confine our attention to the spatial orbital angular momentum or $(\boldsymbol{L})^{a}=\frac{1}{2}\varepsilon^{aij}L^{0ij}$, in which case we can determine that only the initial criteria will contribute yielding
\begin{equation}
    \boldsymbol{L}=\frac{\mathcal{T}^{2}\mathcal{V}}{2\pi^{2}}\frac{\mathcal{E}_{\parallel}^{2}|g\mathcal{B}_{\parallel}|}{\mathcal{B}_{\parallel}^{2}+\mathcal{E}_{\parallel}^{2}}\coth\Bigl(\frac{|\mathcal{B}_{\parallel}|\pi}{|\mathcal{E}_{\parallel}|}\Bigr)e^{-\frac{m^{2}\pi}{|g\mathcal{E}_{\parallel}|}}g^{2}\boldsymbol{z}\times[ \boldsymbol{\mathcal{E}}_z \times \boldsymbol{\mathcal{B}}_z + \boldsymbol{\mathcal{E}}_z \times\boldsymbol{\mathcal{B}}_{\parallel}+\boldsymbol{\mathcal{E}}_{\parallel}\times \boldsymbol{\mathcal{B}}_z ]\,,
\end{equation}
where we have used eq.~\eqref{eq:heuristic_momentum}. We can immediately see that the Schwinger produced pair angular momentum density is proportional to that given by the background fields in eq.~\eqref{eq:F_angular_density} as anticipated. Also, as constructed there is a characteristic exponential quadratic mass suppression. Let us now treat the event averaged scenario in section~\ref{sec:hic}. We can find the complete expression using eq.~\eqref{eq:hic_event} as
\begin{equation}
    \llangle\boldsymbol{L}\rrangle=\frac{|g\mathcal{B}_{\parallel}|}{2\pi^{2}}\frac{\mathcal{E}_{\parallel}^{2}\mathcal{T}^{2}\mathcal{V}}{\mathcal{B}_{\parallel}^{2}+\mathcal{E}_{\parallel}^{2}}\coth\Bigl(\frac{|\mathcal{B}_{\parallel}|\pi}{|\mathcal{E}_{\parallel}|}\Bigr)e^{-\frac{m^{2}\pi}{|g\mathcal{E}_{\parallel}|}}g^{2}\llangle\boldsymbol{l}_{\mathcal{F}}\rrangle\,,
    \label{eq:angmomclasfinal_total}
\end{equation}
where we have made use of eq.~\eqref{eq:F_total_ang_mom}. It is clear the two angular momenta are proportional.
Equation~\eqref{eq:angmomclasfinal_total} indicates how much of the angular momentum of the gluon fields is transferred to those of Schwinger produced pairs.

\section{Nonequilibrium In-In Formalism}
\label{sec:nonequilibrium}

Having seen above the emergence of a net angular momentum from fields possessing an angular momentum in a classically motivated virtual condensate breaking picture, let us calculate the full quantum observable. Since the Schwinger effect is a vacuum unstable phenomenon inherently out-of-equilibrium, one must treat the vacuum state identification appropriately; this can be achieved using an in-in formalism~\cite{etde_6972673}. The in-in formalism is equivalent to a Schwinger-Keldysh (or closed time path or real-time) formalism~\cite{doi:10.1063/1.1703727,PhysRev.124.287}. We make use of a one-loop formulation for an in-in propagator in the Schwinger propertime picture~\cite{etde_6972673}. 

First let us digress on the Schwinger propertime picture for the conventional matrix element, in-out, application. The in-out casual propagator may be cast in Schwinger propertime as
\begin{equation}
    S_\Omega(x,y)=i\langle\textrm{out}|\mathcal{T}\psi(x)\bar{\psi}(y)|\textrm{in}\rangle\langle\textrm{out}|\textrm{in}\rangle^{-1}=(i\slashed{\mathcal{D}}+m)\int_{0}^{\infty}dT\,\mathcal{K}(x,y,T)\,,
    \label{eq:S^c}
\end{equation}
with the kernel given in eq.~\eqref{eq:kernel}. Observables calculated from the in-out propagator are associated with the vacuum polarization quantities~\cite{PhysRevD.78.045017}. We wish to explore specifically quantities related to the vacuum instability, which manifest as out-of equilibrium observables and are captured within the in-in formalism.

The extension to in-in vacuum states in the Schwinger propertime picture has been derived in ref.~\cite{etde_6972673}. There, it was demonstrated the propertime contour may be augmented with a discontinuity about the spacelike electric field eigenvector for the causal propagator as
\begin{align}
    S_{\textrm{in}}(x,y)&=i\langle\textrm{in}|\mathcal{T}\psi(x)\bar{\psi}(y)|\textrm{in}\rangle =(i\slashed{\mathcal{D}}+m)\int_{\textrm{in}}dT\, \mathcal{K}(x,y,T)\,,\\
    \int_{\textrm{in}}dT &\coloneqq \int_{0}^{\infty}dT-\int_{0-i\frac{\pi}{\lambda_{E}}}^{\infty-i\frac{\pi}{\lambda_{E}}}dT-\Theta(z)\lcirclerightint_{-i\frac{\pi}{\lambda_{E}}}dT\,,
    \label{eq:int_in}\\
    \Theta(z)&\coloneqq\theta(n_{\lambda_{E}}^{-\,T}z)\theta[(n_{\lambda_{E}}^{-\,T}z)^{2}-(n_{\lambda_{E}}^{+\,T}z)^{2}]\,.
\end{align}
Here $z\coloneqq x-y$. The final integral in eq.~\eqref{eq:int_in} represents an infinitesimal clockwise closed contour in propertime about $T=-i\pi/\lambda_E$. $\theta$ are the Heaviside theta functions; their argument about the origin, $z=0$, is defined as $S(x,x)=(1/2)[S(x,x+\epsilon)+S(x+\epsilon,x)]$; note that this will lead to $\theta(0)=1/2$.  $n_{\lambda_{E}}^{-}$ and $n_{\lambda_{E}}^{+}$ are respectively the spacelike and timelike combinations of the eigenvectors, which are null, of the electromagnetic field strength, $\mathcal{F}$. We explored their properties in sec.~\ref{sec:wong_sol}; see eq.~\eqref{eq:F_comb_eigenvectors} and thereafter. Notice, furthermore, that the vacuum polarization component, eq.~\eqref{eq:S^c}, is included within the in-in causal propagator.

Since we are concerned with the contribution coming from the nonequilibrium vacuum instability, let us confine our attention to the portion of the propagator without the vacuum polarization as
\begin{equation}
    S_{\textrm{C}}(x,y)\coloneqq S_{\textrm{in}}(x,y)-S_\Omega(x,y)\quad\text{or}\quad \int_{\textrm{C}}\coloneqq \int_{\textrm{in}}-\int^\infty_0\,.
\end{equation}

The kernel, eq.~\eqref{eq:kernel}, can be solved exactly in SU$(2)\times$U$(1)$ Abelian projected homogeneous fields. We demonstrate this calculation in appendix~\ref{sec:kernel}, and for manipulations to follow below it is convenient to gather the results here. The kernel may be cast in diagonal form as $\mathcal{K}(x,y,T)=\textrm{diag}(\mathcal{K}^{+}(x,y,T),\mathcal{K}^{-}(x,y,T))$, with
\begin{align}
    \mathcal{K}^{+}(x,y,T)&=\frac{\lambda_{E}\lambda_{B}\exp\bigl[-im^{2}T+i\varphi(x,y,T)\bigr]}{(4\pi)^{2}\sinh(\lambda_{E}T)\sin(\lambda_{B}T)}\Phi(T)\,,\label{eq:kernel_soln}\\
    \varphi(x,y,T)&=\frac{1}{2}x^T\mathcal{F}y-\frac{1}{4}\Bigl\{\lambda_{E}\coth(\lambda_{E}T)z^TP_{E}z+\lambda_{B}\cot(\lambda_{B}T)z^TP_{B}z\Bigr\}\,,\\
    \Phi(T)&=\cos(\lambda_{B}T)\cosh(\lambda_{E}T)+i\gamma_{5}\textrm{sgn}(I_{\tilde{F}F})\sin(\lambda_{B}T)\sinh(\lambda_{E}T) \nonumber \\
    &-\frac{1}{2}\bigl[\lambda_{B}+i\gamma_{5}\textrm{sgn}(I_{\tilde{F}F})\lambda_{E}\bigr] \times\frac{\mathcal{F}_{\mu\nu}\sigma^{\mu\nu}}{\lambda_{B}^{2}+\lambda_{E}^{2}}\Bigl[i\sin(\lambda_{B}T)\cosh(\lambda_{E}T)\nonumber \\
    &+\gamma_{5}\textrm{sgn}(I_{\tilde{F}F})\cos(\lambda_{B}T)\sinh(\lambda_{E}T)\Bigr]\,.
    \label{eq:spin_factor_final}
\end{align}
$\mathcal{K}^{-}(x,y,T)$ can be had with the replacement $g\rightarrow-g$. 

\subsection{Spin Angular Momentum}
\label{sec:spin_ang_mom}

Let us first calculate the angular momentum coming from spin. The spin angular momentum at operator level is $(1/2)\bar{\psi}\gamma^{\mu}\sigma^{\nu\sigma}\psi$, with nonequilibrium expectation value
\begin{equation}
    S^{\mu\nu\sigma}=\frac{i}{2}\lim_{x\rightarrow y} \mathrm{tr}_c  \mathrm{tr}_\gamma [\gamma^{\mu}\sigma^{\nu\sigma}S_{\textrm{C}}(x,y)]\,.
\end{equation}
Let us restrict our attention to the case of spatial spin angular momentum; this is nothing but the axial current since $\gamma^{0}\sigma^{ij} =-\varepsilon^{kij}\gamma_{k}\gamma^{5}$. Calculations here then are similar to those discussed in ref.~\cite{doi:10.1142/S0217751X2030015X}. Then, applying the kernel solution given in eq.~\eqref{eq:kernel_soln} one can find that
\begin{equation}
    S^{0ij}=-\frac{1}{2}\lim_{x\rightarrow y} \mathrm{tr}_\gamma \varepsilon^{kij}\gamma_{k}\gamma^{5}[\slashed{\partial}_{x}\Theta(z)]\lcirclerightint_{-i\frac{\pi}{\lambda_{E}}}dT\,\Bigl\{\mathcal{K}^{+}(x,y,T)+\mathcal{K}^{-}(x,y,T)\Bigr\}\,.
\end{equation}
Let us confine our attention to just the $\mathcal{K}^{+}$ part; we refer to the part of the above with just the $g\rightarrow +g$ component as $S^{+\,0ij}$. And we use this notation for similar calculations from here on. About the closed contour contributions from $\varphi$ will not contribute in the $z\rightarrow 0$ limit~\cite{doi:10.1142/S0217751X2030015X}. We have
\begin{equation}
    S^{+\,0ij} =-\frac{1}{2}\lim_{x\rightarrow y}\lcirclerightint_{-i\frac{\pi}{\lambda_{E}}}dT\,\frac{\lambda_{E}\lambda_{B}\exp(-im^{2}T)}{(4\pi)^{2}\sinh(\lambda_{E}T)\sin(\lambda_{B}T)}\mathrm{tr}_\gamma [\varepsilon^{kij}\gamma_{k}\gamma^{5}\slashed{\partial}_{x}\Theta(z)\Phi(T)]\,.
\end{equation}
Since pieces without a singularity will vanish after taking the integral about the closed contour we find for the quantity in the Dirac trace above the following:
\begin{equation}
    -\frac{4i}{(\lambda_{B}^{2}+\lambda_{E}^{2})}\varepsilon^{kij}\bigl[\lambda_{B}\widetilde{F}^{k\nu}+\textrm{sgn}(I_{\tilde{F}F})\lambda_{E}\mathcal{F}^{k\nu}\bigr]\partial_{x\,\nu}\Theta(z)\sin(\lambda_{B}T)\cosh(\lambda_{E}T)\,.
\end{equation}

One can see, here, the appearance of a delta function singularity, $\delta(n_{\lambda_{E}}^{-\,T}z)$, caused by $\partial_{x\,\nu}\Theta(z)$. This formally divergent contribution is associated with the real-time of the system~\cite{Nikishov:1969tt}. It arises in homogeneous electric fields through the appearance of a discontinuity in the propertime integration~\cite{doi:10.1063/1.532451}, however divergent expectation values can be connected to convergent equivalents in a switch-on electric field for large switch-on times~\cite{PhysRevD.78.045017}. The real-time manifestation is a consequence of a cutoff in the canonical momentum of the system as
\begin{equation}
    \lim_{n_{\lambda_{E}}^{-\,T}z\rightarrow0}\delta(n_{\lambda_{E}}^{-\,T}z)=\lim_{n_{\lambda_{E}}^{-\,T}z\rightarrow0}\int_{-\lambda_E\mathcal{T}/2}^{\lambda_E\mathcal{T}/2}\frac{dp}{2\pi}e^{ip(n_{\lambda_{E}}^{-\,T}z)}=\frac{\lambda_{E}\mathcal{T}}{2\pi}\,.
    \label{eq:delta_singularity}
\end{equation}
Solutions to the Dirac equation in a homogeneous electric field give rise to an admixture of particle and antiparticle states at times corresponding to the cutoff above, and can thus be regarded as the time to produce pairs~\cite{Nikishov_long}. For further discussions on the interpretation also see ref.~\cite{TANJI20091691}.

The spin angular momentum becomes
\begin{equation}
    S^{+\,0ij}=\frac{1}{\lambda_{B}^{2}+\lambda_{E}^{2}}\varepsilon^{kij}\bigl[\bigl(\lambda_{B}\widetilde{F}+\textrm{sgn}(I_{\tilde{F}F})\lambda_{E}\mathcal{F}\bigr)n_{\lambda_{E}}^{-}\bigr]^k\frac{\lambda_{E}\lambda_{B}\mathcal{T}}{4\pi{}^{2}}e^{-\frac{\pi m^{2}}{\lambda_{E}}}\,.
\end{equation}
Finally using eqs.~\eqref{eq:Fn} and \eqref{eq:tildeFn}, we arrive at
\begin{equation}
    S^{0ij}=-\varepsilon^{kij}\frac{I_{\tilde{F}F}\mathcal{T}}{4\pi{}^{2}}e^{-\frac{\pi m^{2}}{\lambda_{E}}}n_{\lambda_{E}}^{+\,k}+[g\rightarrow-g]\,.
\end{equation}
In the HIC setup in section \ref{sec:hic}, $I_{\tilde{F}F}$ is invariant under $g\to -g$, $I_{\tilde{F}F}\approx g^2\boldsymbol{\mathcal{E}}_{\parallel}\cdot\boldsymbol{\mathcal{B}}_{\parallel}$, so that we can write $S^{0ij}$ as 
\begin{equation}
    S^{0ij}=-\varepsilon^{kij}\frac{
    g^2\boldsymbol{\mathcal{E}}_{\parallel}\cdot\boldsymbol{\mathcal{B}}_{\parallel}
\mathcal{T}}{4\pi{}^{2}}e^{-\frac{\pi m^{2}}{\lambda_{E}}}
    (n_{\lambda_{E}}^{+\,k}|_{g\rightarrow g}+
    n_{\lambda_{E}}^{+\,k}|_{g\rightarrow -g})\,.
    \label{eq:spinangmom}
\end{equation}
Although the expression of $n_{\lambda_{E}}^{+\,k}$ is generally complicated, 
thanks to the parity and time reversal properties of $S^{0ij}$, $(n_{\lambda_{E}}^{+\,k}|_{g\rightarrow g}+
    n_{\lambda_{E}}^{+\,k}|_{g\rightarrow -g})$ should be proportional to $g\boldsymbol{\mathcal{E}}\times g\boldsymbol{\mathcal{B}}=g^2( \boldsymbol{\mathcal{E}}_z \times \boldsymbol{\mathcal{B}}_z + \boldsymbol{\mathcal{E}}_z \times\boldsymbol{\mathcal{B}}_{\parallel}+\boldsymbol{\mathcal{E}}_{\parallel}\times \boldsymbol{\mathcal{B}}_z )$.
After averaging over $\pm \mathcal{E}_\parallel$ and $\pm \mathcal{B}_\parallel$, 
we obtain that $\llangle S^{0ij} \rrangle =0$, as anticipated in the heuristic picture.

Let us gain a better grasp of the expression not using the fields depicted in section~\ref{sec:hic}, but rather those of parallel electric and magnetic fields in the $\hat{\boldsymbol{x}}_3$ direction, with electric field given as $\mathcal{E}\hat{\boldsymbol{x}}_3$. For such a field the electromagnetic eigenvectors take on a simple form with $n_E^{+\mu}=g^{0\mu} - (g\mathcal{E}/|g\mathcal{E}|)g^{3 \mu}$ and $n_E^{-\mu}=g^{0\mu} + (g\mathcal{E}/|g\mathcal{E}|)g^{3 \mu}$. Then the combined eigenvectors read simply $n_{\lambda_E}^{+\mu}=g^{0\mu}$ and $n_{\lambda_E}^{-\mu}=(g\mathcal{E}/|g\mathcal{E}|)g^{3 \mu}$. One can then see that a non-vanishing time component of $n_{\lambda_E}^{+0}$ persists even when summing over $g\rightarrow -g$; this is the chiral density: $(I_{\tilde{F}F} \mathcal{T}/ 2\pi^2)\exp(-\pi m^{2} / \lambda_{E})g^{0 \mu}$, thus confirming the heuristic expression given in eq.~\eqref{eq:p_0_total}. Such a chiral density has been shown to be in agreement with the axial anomaly in the massless limit with an analogous calculation of the pseudoscalar condensate and Chern-Simons term making up the axial Ward identity~\cite{PhysRevLett.121.261602,doi:10.1142/S0217751X2030015X}. 

One may go through an analogous calculation of the vector current, that is $j^{\mu}=ie \lim_{x\rightarrow y}  \mathrm{tr}_c  \mathrm{tr}_\gamma [\gamma^{\mu}S_{\textrm{C}}(x,y)]$,
to find the conduction out-of equilibrium current associated with Schwinger pair production~\cite{etde_6972673} as
\begin{equation}
    j^{\mu}=e\frac{\lambda_{E}\lambda_{B}\mathcal{T}}{2\pi^{2}}e^{-\frac{\pi m^{2}}{\lambda_{E}}}\coth\bigl(\frac{\pi\lambda_{B}}{\lambda_{E}}\bigr)\,n_{\lambda_{E}}^{-\,\mu}+[g\rightarrow-g]\,.
    \label{eq:inin_current}
\end{equation}
Let us again make use of the parallel electric and magnetic field described above. Then it can be seen that due to the odd in $g$ factor in $n_{\lambda_{E}}^{-\,\mu}$, the conduction current in SU$(2)\times$U$(1)$ will vanish~\cite{PhysRevD.92.125012}, confirming the heuristic picture quantity found in eq.~\eqref{eq:current_heuristic}.

\subsection{Orbital Angular Momentum}
\label{sec:orbital_ang_mom}

The other addition to the angular momentum comes from the orbital angular momentum.
Since the spin angular momentum vanishes in our setup,
the total angular momentum coincides with the orbital one. Let us here consider the total angular momentum, whose density is given as
\begin{equation}
l^{\mu\nu\sigma}=x^{\nu}\mathbb{T}_{\textrm{C}}^{\mu\sigma}-x^{\sigma}\mathbb{T}_{\textrm{C}}^{\mu\nu}\,,
\label{eq:l_def}
\end{equation}
where the fully symmetric stress energy tensor reads
\begin{equation}
\mathbb{T}^{\mu\nu}=\frac{1}{4}\langle\textrm{in}|\bar{\psi}\gamma^{(\mu}i\mathcal{D}^{\nu)}\psi-\bar{\psi}i\overleftarrow{\mathcal{D}}^{(\nu}\gamma^{\mu)}\psi|\textrm{in}\rangle-g^{\mu\nu}\langle\textrm{in}|\bar{\psi}(i\slashed{\mathcal{D}}-m)\psi |\textrm{in}\rangle\,.
\end{equation}
Here $A_{(\mu}B_{\nu)}=A_{\mu}B_{\nu}+A_{\nu}B_{\mu}$, and we have assumed by construction of the in-in propagator an implicit averaging over Dirac operator order in the coincidence, $x\rightarrow y$, limit. 
The second term can be dropped by using the Dirac equation.
And as before let us break up vacuum polarization and vacuum instability parts as $\mathbb{T}^{\mu\nu}=\mathbb{T}_{\textrm{C}}^{\mu\nu}+\mathbb{T}_{\Omega}^{\mu\nu}$. It is simpler to explicitly treat 
\begin{equation}
T^{\mu\nu}=\langle\textrm{in}|\bar{\psi}\gamma^{\mu}i\mathcal{D}^{\nu}\psi|\textrm{in}\rangle\,,
\end{equation}
then one can find $\mathbb{T}^{\mu\nu}=(1/4)(T^{(\mu\nu)}+[T^{(\mu\nu)}]^{*})$. Also $T^{\mu\nu}=T_\mathrm{C}^{\mu\nu}+T_{\Omega}^{\mu\nu}$, where
\begin{equation}
T_{\textrm{C}}^{\mu\nu}=-\lim_{x\rightarrow y} \mathrm{tr}_c  \mathrm{tr}_\gamma [\gamma^{\mu}\mathcal{D}_{x}^{\nu}S_{\textrm{C}}(x,y)]\,,\quad
T_{\Omega}^{\mu\nu}=-\lim_{x\rightarrow y} \mathrm{tr}_c  \mathrm{tr}_\gamma [\gamma^{\mu}\mathcal{D}_{x}^{\nu}S_{\Omega}(x,y)]\,.
\label{eq:OmegaC}
\end{equation}
Let us write as before again splitting up the $\pm g$ parts 
\begin{equation}
T^{\mu\nu}=-\lim_{x\rightarrow y} \mathrm{tr}_\gamma \gamma^{\mu}\biggl\{\mathcal{D}_{x}^{\nu\,+}(i\slashed{\mathcal{D}}_{x}^{+}+m)\int_{\textrm{in}}dT\,\mathcal{K}^{+}(x,y,T)+\mathcal{D}_{x}^{\nu\,-}(i\slashed{\mathcal{D}}_{x}^{-}+m)\int_{\textrm{in}}dT\,\mathcal{K}^{-}(x,y,T)\biggr\}\,.
\end{equation}
We may drop the terms proportional to mass because $\mathcal{K}$ contains an even number of $\gamma^{\mu}$, and thus the trace vanishes.
It is convenient for us in the following to recast the in-in propertime integral as
\begin{equation}
\int_{\textrm{in}}dT\,\mathcal{K}^{+}(x,y,T)=\biggl\{\int_{0}^{\infty}dT-\int_{0-i\frac{\pi}{\lambda_{E}}}^{\infty-i\frac{\pi}{\lambda_{E}}}dT-\theta(n_{\lambda_{E}}^{-\,T}z) \curvearrowright\mathclap{\int_{-i\frac{\pi}{\lambda_{E}}}} \quad dT\biggr\}\mathcal{K}^{+}(x,y,T)\,;
\end{equation}
the integral on the right denotes a semicircle contour from $T=-i\pi/\lambda_{E}-0$ to $T=-i\pi/\lambda_{E}+0$  going over the pole at $T=-i\pi/\lambda_{E}$. The essential singularities at $T=-in\pi/\lambda_{E}$ in $\mathcal{K}^+$ may be broken into upper and lower semicircle contours each of which is finite within restricted light cone electric field variables, namely for either $\theta(\pm[(n_{\lambda_{E}}^{-\,T}z)^{2}-(n_{\lambda_{E}}^{+\,T}z)^{2}])$; see ref.~\cite{etde_6972673} for further discussions. Let us show the two covariant derivatives together commute through the Heaviside function. We first write
\begin{equation}
[\mathcal{D}^{\mu +}\mathcal{D}^{\nu +},\theta(n_{\lambda_{E}}^{-\,T}z)]\mathcal{K}^{+} =\Bigl[n_{\lambda_{E}}^{-\nu}(\partial^{\mu}\delta(n_{\lambda_{E}}^{-\,T}z))+n_{\lambda_{E}}^{-(\mu}\delta(n_{\lambda_{E}}^{-\,T}z)\mathcal{D}^{\nu)}\Bigr]\mathcal{K}^{+}\,.
\end{equation}
The delta functions in the above are even functions in $z$. Since $i\mathcal{D}_{x}^{+}\mathcal{K}^{+}(x,y,T)=\frac{1}{2}[\mathcal{F}+\coth(\mathcal{F}T)\mathcal{F}]z\mathcal{K}^{+}(x,y,T)$ is odd in $z$, once we take the coincidence limit, (i.e., $z\rightarrow0$, where we average over both $z\rightarrow\pm\epsilon$ for $\epsilon$ small), such terms will vanish. Let us remark that even though there are divergences after taking the proper time integral, the fact that the final expression is still odd in $z$ shows the translational invariance and hence why such terms disappear~\cite{10.1007/3-540-15653-4_3}. One may show in an analogous fashion using
\begin{equation}
\partial_{x}^{\mu}\delta(n_{\lambda_{E}}^{-\,T}z)=\partial_{x}^{\mu} \int_{-\lambda_E\mathcal{T}/2}^{\lambda_E\mathcal{T}/2}\frac{dp}{2\pi}e^{ipn_{\lambda_{E}}^{-\,T}z}=n_{\lambda_{E}}^{-\,\mu}\int_{-\lambda_E\mathcal{T}/2}^{\lambda_E\mathcal{T}/2}\frac{dp}{2\pi}ipe^{ip(n_{\lambda_{E}}^{-\,T}z)}\,,
\end{equation}
that the corresponding term there too vanishes in the coincidence limit, and we find that
\begin{equation}
T^{\mu\nu+}=-\lim_{x\rightarrow y} \mathrm{tr}_\gamma \gamma^{\mu}\int_{\textrm{in}}dT\,\mathcal{D}_{x}^{\nu\,+}i\slashed{\mathcal{D}}_{x}^{+}\mathcal{K}^{+}(x,y,T)\,.
\end{equation}

The covariant derivatives acting on the kernel read
\begin{equation}
i\mathcal{D}_{x}^{\nu\,+}i\mathcal{D}_{x}^{\sigma\,+}\mathcal{K}^{+}=\Biggl\{\frac{i}{2}[\mathcal{F}+\coth(\mathcal{F}T)\mathcal{F}]^{\sigma\nu}+\frac{1}{4}[\mathcal{F}z+\coth(\mathcal{F}T)\mathcal{F}z]^{\sigma}[\mathcal{F}z+\coth(\mathcal{F}T)\mathcal{F}z]^{\nu}\Biggr\}\mathcal{K}^{+}\,.
\end{equation}
Let us next evaluate the Dirac trace. This is $\mathrm{tr}_\gamma \gamma^{\mu}\gamma^{\sigma}\Phi(T)$, with spin factor given by eq.~\eqref{eq:spin_factor_final}. Making use of the following identities,
\begin{align}
 & \mathrm{tr}_\gamma \frac{1}{2(\lambda_{B}^{2}+\lambda_{E}^{2})}\Bigl\{\gamma^{\mu}\gamma^{\sigma}(\lambda_{B}+i\gamma_{5}\textrm{sgn}(I_{\tilde{F}F})\lambda_{E})\mathcal{F}_{\mu'\nu'}\sigma^{\mu'\nu'}\gamma_{5}\Bigr\}=\frac{4}{\lambda_{B}}\bigl[P_{E}\widetilde{\mathcal{F}}\bigr]^{\mu\sigma}\,,\\
 & \mathrm{tr}_\gamma \frac{i}{2(\lambda_{B}^{2}+\lambda_{E}^{2})}\Bigl\{\gamma^{\mu}\gamma^{\sigma}(\lambda_{B}+i\gamma_{5}\textrm{sgn}(I_{\tilde{F}F})\lambda_{E})\mathcal{F}_{\mu'\nu'}\sigma^{\mu'\nu'}\Bigr\}=-4\frac{\lambda_{B}}{I_{\tilde{F}F}}\bigl[P_{B}\widetilde{\mathcal{F}}\bigr]^{\mu\sigma}\,,
\end{align}
we can find that
\begin{align}
& \mathrm{tr}_\gamma \gamma^{\mu}\gamma^{\sigma}\Phi(T)=4\Biggl\{ g^{\mu\sigma}\cos(\lambda_{B}T)\cosh(\lambda_{E}T)\nonumber\\&+\frac{\lambda_{B}}{I_{\tilde{F}F}}\bigl[P_{B}\widetilde{\mathcal{F}}\bigr]^{\mu\sigma}\sin(\lambda_{B}T)\cosh(\lambda_{E}T)-\frac{\lambda_{E}}{I_{\tilde{F}F}}[P_{E}\widetilde{\mathcal{F}}]^{\mu\sigma}\cos(\lambda_{B}T)\sinh(\lambda_{E}T)\Biggr\}\,.
\end{align}
And the traced kernel becomes
\begin{align}
T^{\mu\nu\,+} & =i\lim_{x\rightarrow y}\int_{\textrm{in}}dT\,\frac{\lambda_{E}\lambda_{B}e^{-im^{2}T+i\varphi}}{4\pi^{2}}\Biggl\{\frac{i}{2}W(T)^{\mu\nu}+\frac{1}{4}[W(T)z]^{\mu}[\mathcal{F}z+\coth(\mathcal{F}T)\mathcal{F}z]^{\nu}\Biggr\}\,,
\end{align}
where 
\begin{equation}
W(T)=\coth(\lambda_{E}T)\lambda_{B}P_{B}-\cot(\lambda_{B}T)\lambda_{E}P_{E}+\cot(\lambda_{B}T)\coth(\lambda_{E}T)\coth(\mathcal{F}T)\mathcal{F}\,.
\end{equation}

Having written the traced kernel in a compact form, let us address the various integrals. Notice contributions close to the singularities at $T=-in\pi/\lambda_{E}$ are formally divergent; as anticipated earlier, this divergence in momentum indicates a total time of the electric field of the system. Therefore let us approximate the integrals by expanding about such points. For the in-in propagator only $n=0,1$ will contribute; one can easily see this by closing the contour about $\int_{0}^{\infty}dT-\int_{0-i\pi/\lambda_{E}}^{\infty-i\pi/\lambda_{E}}dT$. Furthermore let us now break up the energy momentum tensor as instructed in eq.~\eqref{eq:OmegaC}, then we find
\begin{align}
T_{\Omega}^{\mu\nu\,+}&\approx i\lim_{x\rightarrow y}\int_{0}^{\infty}dT\frac{e^{-im^{2}T-i\frac{1}{4T}z^{T}z}}{4\pi^{2}}\biggl\{\frac{i}{2}\frac{g^{\mu\nu}}{T^{3}}+\frac{z^{\mu}z^{\nu}}{4T^4}\biggr\}\,,\\
T_{\textrm{C}}^{\mu\nu\,+}&=-i\lim_{x\rightarrow y}\biggl[\int_{0}^{\infty}dT+\theta(n_{\lambda_{E}}^{-\,T}z) \;\;\mathclap{\curvearrowright}\mathclap{\int_0} \quad dT\biggr]\frac{\lambda_{E}\lambda_{B}e^{-im^{2}T-\frac{m^{2}\pi}{\lambda_{E}}+i\varphi\bigl(T-i\frac{\pi}{\lambda_{E}}\bigr)}}{4\pi^{2}}\nonumber \\
&\times\Biggl\{\frac{i}{2}W^{\mu\nu}\Bigl(T-i\frac{\pi}{\lambda_{E}}\Bigr)+\frac{1}{4}\Bigl[W\Bigl(T-i\frac{\pi}{\lambda_{E}}\Bigr)z\Bigr]^{\mu}\Bigl[\coth\Bigl(\mathcal{F}\Bigl(T-i\frac{\pi}{\lambda_{E}}\Bigr)\Bigr)\mathcal{F}z\Bigr]^{\nu}\Biggr\}\\
&\approx-i\lim_{x\rightarrow y}\biggl[\int_{0}^{\infty}dT+\theta(n_{\lambda_{E}}^{-\,T}z) \;\;\mathclap{\curvearrowright}\mathclap{\int_0} \quad dT\biggr]\frac{\lambda_{E}\lambda_{B}e^{-im^{2}T-\frac{m^{2}\pi}{\lambda_{E}}-\frac{i}{4T}z^{T}P_{E}z}}{4\pi^{2}}\nonumber \\
 & \quad\times\Biggl\{-\frac{1}{2}\coth\Bigl(\frac{\pi\lambda_{B}}{\lambda_{E}}\Bigr)\frac{P_{E}^{\mu\nu}}{\lambda_{E}T^{2}}+\frac{i}{4}\Bigl[\coth\Bigl(\frac{\pi\lambda_{B}}{\lambda_{E}}\Bigr)\frac{P_{E}z}{\lambda_{E}T^{2}}\Bigr]^{\mu}\Bigl[\frac{P_{E}z}{T}\Bigr]^{\nu}\Biggr\}\,.
\end{align}
The integral on the right denotes a similar semicircle contour as before but about the origin $T=0$. 

We can see that ultimately for $T_{\Omega}^{\mu\nu+}$ there will be no field dependence as expected, and will be unrelated to Schwinger pair production. The term predicts the vacuum polarization energy-momentum quantities. There are UV divergences; and there are formal divergences in $T_{\text{C}}^{\mu\nu+}$ as well. Let us simply use a propertime UV cutoff of $\Lambda^{-2}$ to illustrate briefly such divergences in $T_{\Omega}^{\mu\nu+}$. Then one can find that the vacuum polarization energy-momentum tensor goes like $T_{\Omega}^{\mu\nu+}=(\Lambda^{4}/16\pi^{2})g^{\mu\nu}+\mathcal{O}(\Lambda^{2})$~\cite{PhysRevD.39.3478}. From hereafter we will treat quantities associated with a Schwinger pair production conduction current described in $T_{\text{C}}^{\mu\nu+}$. Divergences there, however and as we encountered previously, predict a real-time like dependence and furthermore the coordinates associated with magnetic degrees of freedom are not present, therefore we treat such singular structures carefully, and elect to use a UV cutoff in the canonical momentum integral after Fourier transform. It is convenient to introduce a shorthand notation such that the two coordinate degrees of freedom associated with the electric field may be written as $n_{\lambda_{E}}^{+\,T}z\coloneqq t_{E}$ and $n_{\lambda_{E}}^{-\,T}z\coloneqq z_{E}$, then we label the integrals as
\begin{equation}
f_{n}(z)\coloneqq \int_{0}^{\infty}\frac{dT}{T^n}e^{-im^{2}T-i\frac{1}{4T}z^{T}P_{E}z}=\int_{0}^{\infty}\frac{dT}{T^n}\,e^{-im^{2}T-i\frac{1}{4T}(t_{E}^{2}-z_{E}^{2})}\,.
\end{equation}
However, one need only evaluate $f_{1}(z)$, and use the fact that $\partial^{\mu}\partial^{\nu}f_{1}(z)=-(i/2)P_{E}^{\mu\nu}f_{2}(z)-(1/4)(P_{E}z)^{\mu}(P_{E}z)^{\nu} f_{3}(z)$. $f_{1}(z)$ in fact resembles a two dimensional solution to a Klein-Gordon equation, and it proves convenient to replace the propertime integral with one over canonical momentum. Furthermore the formally divergent large momenta indicate real-time dependence. Therefore, let us take the Fourier then the inverse Fourier transforms; see~\cite{greiner2008quantum} for a similar representation of the propagator in 3+1-dimensions:
\begin{align}
f_{1}(p)&=\int_{0}^{\infty}\frac{dT}{T}\int dt_{E}dz_{E}e^{ip_{t_{E}}t_{E}-ip_{z_{E}}z_{E}}e^{-im^{2}T-i\frac{1}{4T}(t_{E}^{2}-z_{E}^{2})}=\frac{4\pi i}{p_{t_{E}}^{2}-p_{z_{E}}^{2}-m^{2}+i\epsilon}\,,\\
f_{1}(z)&=\int\frac{dp_{t_{E}}}{2\pi}\frac{dp_{z_{E}}}{2\pi}e^{-ip_{t_{E}}t_{E}+ip_{z_{E}}z_{E}}f_1(p)=\int dp_{z_{E}}\frac{e^{-i\sqrt{p_{z_{E}}^{2}+m^{2}}|t_{E}|+ip_{z_{E}}z_{E}}}{\sqrt{p_{z_{E}}^{2}+m^{2}}}\,.
\end{align}
Then taking the derivatives we have
\begin{align}
&\partial^{\mu}\partial^{\nu}f_{1}(z)=\int dp_{z_{E}}\Biggl\{-n_{\lambda_{E}}^{+\mu}n_{\lambda_{E}}^{+\nu}\sqrt{p_{z_{E}}^{2}+m^{2}}-n_{\lambda_{E}}^{-\nu}n_{\lambda_{E}}^{-\mu} \frac{p_{z_{E}}^{2}}{\sqrt{p_{z_{E}}^{2}+m^{2}}}\nonumber \\ &+(n_{\lambda_{E}}^{+\mu}n_{\lambda_{E}}^{-\nu}+n_{\lambda_{E}}^{+\nu}n_{\lambda_{E}}^{-\mu})\textrm{sgn}(n_{\lambda_{E}}^{+\,T}z)p_{z_{E}}-2 in_{\lambda_{E}}^{+\nu}n_{\lambda_{E}}^{+\mu}\delta(n_{\lambda_{E}}^{+\,T}z)\Biggr\}e^{-i\sqrt{p_{z_{E}}^{2}+m^{2}}|n_{\lambda_{E}}^{+\,T}z|+ip_{z_{E}}n_{\lambda_{E}}^{-\,T}z}\,.
\label{eq:f_1_derivative}
\end{align}
To evaluate the coincidence limit, we use conventions as outlined in section~\ref{sec:intro}. Also, as before we introduce a cutoff in the canonical momentum for physical background fields; this is $\int_{-\lambda_{E}\mathcal{T}/2}^{\lambda_{E}\mathcal{T}/2}dp_{z_{E}}$,
\begin{equation}
\lim_{x\rightarrow y}\partial^{\mu}\partial^{\nu}f_{1}(z) =-(2n_{\lambda_{E}}^{-\nu}n_{\lambda_{E}}^{-\mu}+P_E^{\mu\nu})\frac{\lambda_{E}^{2}\mathcal{T}^{2}}{4}-P_{E}^{\mu\nu}m^{2}\ln\Bigl(\frac{\lambda_{E}\mathcal{T}}{m}\Bigr)-\lim_{x\rightarrow y}4\pi in_{\lambda_{E}}^{+\nu}n_{\lambda_{E}}^{+\mu}\delta(n_{\lambda_{E}}^{+\,T}z)\delta(n_{\lambda_{E}}^{-\,T}z)\,.
\label{eq:f_1_final}
\end{equation}
We see there is an imaginary part in the tensor (where for later comparison is simpler to leave without the cutoff), which we will show is cancelled with the contributions coming from the singularity. Let us show that the integral about the singularity is a Heaviside theta function argument
\begin{align}
&\;\;\mathclap{\curvearrowright}\mathclap{\int_0} \quad\frac{dT}{T}e^{-im^{2}T-i\frac{1}{4T}z^{T}P_{E}z}\approx\;\;\mathclap{\curvearrowright}\mathclap{\int_0} \quad\frac{dT}{T}e^{-i\frac{1}{4T}z^{T}P_{E}z}\notag\\&=-\int_{-\infty}^{\infty}\frac{dT'}{T'-i\epsilon}e^{-i\frac{T'}{4}z^{T}P_{E}z}=-2\pi i\theta\Bigl(-\frac{1}{4}z^{T}P_{E}z\Bigr)\,.
\end{align}
Taking the derivatives and limit we can find that
\begin{equation}
\lim_{x\rightarrow y}\theta(n_{\lambda_{E}}^{-\,T}z)\partial_{x}^{\mu}\partial_{x}^{\nu}\theta\Bigl(-\frac{1}{4}z^{T}P_{E}z\Bigr)=-\lim_{x\rightarrow y}P_{E}^{\mu\nu}\delta(n_{\lambda_{E}}^{-\,T}z)\delta(n_{\lambda_{E}}^{+\,T}z)\,.
\end{equation}
In light of the above and eq.~\eqref{eq:f_1_final}, we can confirm that upon taking the trace in the energy-momentum tensor, i.e., $T_{\text{C}\,\mu}^{\mu+}$, no imaginary part will reside, which must be the case due to the Hermiticity construction of the in-in formalism. Furthermore, imaginary pieces will vanish in the fully symmetric definition and therefore we finally have
\begin{equation}
\mathbb{T}_{\textrm{C}}^{\mu\nu+}=\coth\Bigl(\frac{\pi\lambda_{B}}{\lambda_{E}}\Bigr)\frac{\lambda_{B}}{4\pi^{2}}e^{-\frac{\pi m^{2}}{\lambda_{E}}}\Biggl\{(P_{E}^{\mu\nu}+2n_{\lambda_{E}}^{-\mu}n_{\lambda_{E}}^{-\nu})\frac{\lambda_{E}^{2}\mathcal{T}^{2}}{4}+P_{E}^{\mu\nu}m^{2}\ln\Bigl(\frac{\lambda_{E}\mathcal{T}}{m}\Bigr)\Biggr\}\,.
\label{eq:T_final}
\end{equation}

Let us pause at this point to highlight the fact that the trace of the energy-momentum tensor is related to the chiral condensate.
First, however, let us mention that since we do not treat quantum higher-loop corrections of the electromagnetic field, our analysis corresponds to a tree level calculation, and we do not see a trace anomaly of the energy momentum tensor. Then, one can easily show that $\mathbb{T}_{\;\mu}^{\mu}=m\Sigma$, where $\Sigma\coloneqq\langle\textrm{in}|\bar{\psi}\psi|\textrm{in}\rangle$, and that $\Sigma_{\textrm{C}}^{+}\sim m\ln(\lambda_{E}\mathcal{T}/m)$. And this serves also as a check of the above; one would expect a term proportional to the mass simply by looking at the fact that $\Sigma^{+}=im\lim_{x\rightarrow y} \mathrm{tr}_\gamma \int_{\textrm{in}}dT\,\mathcal{K}^{+}(x,y,T)$. Moreover, in the same way one would expect the other term, of $\mathcal{O}(\mathcal{T}^{2})$, be traceless and not contribute to the chiral condensate, but should contribute to the momentum and hence angular momentum of the system.

As we encountered before with the axial vector current, eq.~\eqref{eq:spinangmom}, and vector current, eq.~\eqref{eq:inin_current}, notice there are parts of the energy-momentum tensor that depend on the space-like eigenvector, $n_{\lambda_{E}}^{-\mu}$; however, similar as we had reasoned for the two currents, one could find a special (center of mass) frame in which $n_{\lambda_{E}}^{-0}$ would vanish. Alternatively, using eq.~\eqref{eq:proj_operators}, one may find in another special frame that the time-like eigenvector $n_{\lambda_{E}}^{+0}$ vanishes, which would take $P_E^{0i}\rightarrow-P_E^{0i}$ in some new frame in eq.~\eqref{eq:T_final}, indicating a shift in momentum. Let us treat the former transformation, and we can then evaluate the angular momentum of the system, eq.~\eqref{eq:l_def}, as
\begin{equation}
l^{0ij}=\frac{\lambda_{B}}{(4\pi)^{2}}e^{-\frac{\pi m^{2}}{\lambda_{E}}}(\lambda_{E}\mathcal{T})^{2}\coth\Bigl(\frac{\pi\lambda_{B}}{\lambda_{E}}\Bigr)[x^{i}P_{E}^{0j}-x^{j}P_{E}^{0i}]+[g\rightarrow-g]+\mathcal{O}(\ln \mathcal{T})\,.
\end{equation}
Using $(\boldsymbol{l})^{a}=\frac{1}{2}\varepsilon^{aij}l^{0ij}$ we can find
\begin{equation}
\boldsymbol{l}\approx\frac{\lambda_{B}}{(4\pi)^{2}}e^{-\frac{\pi m^{2}}{\lambda_{E}}}\coth\Bigl(\frac{\pi\lambda_{B}}{\lambda_{E}}\Bigr)\frac{(\lambda_{E}\mathcal{T})^{2}}{\lambda_{E}^{2}+\lambda_{B}^{2}}\boldsymbol{x}\times(g\boldsymbol{\mathcal{E}}\times(e\boldsymbol{B}+g\boldsymbol{\mathcal{B}}))+[g\rightarrow-g]\,.
\end{equation}
Finally using the scenario depicted in section~\ref{sec:heuristic}, we can determine the event averaged angular momentum as
\begin{equation}
    \llangle\boldsymbol{l}\rrangle=\frac{|g\mathcal{B}_{\parallel}|}{8\pi^{2}}e^{-\frac{\pi m^{2}}{|g\mathcal{E}_{\parallel}|}}\coth\Bigl(\frac{\pi|\mathcal{B}_{\parallel}|}{|\mathcal{E}_{\parallel}|}\Bigr)\frac{\mathcal{E}_{\parallel}^{2}\mathcal{T}^{2}}{\mathcal{E}_{\parallel}^{2}+\mathcal{B}_{\parallel}^{2}}g^{2}\llangle\boldsymbol{l}_{\mathcal{F}}\rrangle\,,
    \label{eq:l_final}
\end{equation}
which is proportional to the quantity found using entirely classical and heuristic arguments in eq.~\eqref{eq:angmomclasfinal_total}. This then confirms that for fields which possess a net angular momentum, produced Schwinger pairs too should be proportional to the angular momentum. 
Let us mention that there is Abelian magnetic field dependence in the angular momentum density before the averaging over events,  however after averaging over it disappears as a product of the fields depicted in section~\ref{sec:hic}.

Let us point out that, however, there is a factor of 4 discrepancy between the one-loop quantum computation of eq.~\eqref{eq:l_final} and the heuristic computation of eq.~\eqref{eq:angmomclasfinal_total}. This discrepancy stems from a limitation of the heuristic picture in summing over the momentum. In the heuristic picture each pair of particles is reasoned to occur with (small) probability governed by the Schwinger non-persistence criteria, eq.~\eqref{eq:nonpersistence}, by means of a multiplicative factor. The sum over momenta in the factor goes as $2\int^{\lambda_E \mathcal{T}/2}_0 dp=\lambda_E \mathcal{T}$. And one would pick up another factor for the momentum $p\sim \lambda_E \mathcal{T}$ such that the momentum of the heuristically motivated stress-energy tensor would go as $(\lambda_E \mathcal{T})^2$. However, for the momentum associated with tensor, we must have the linear term in momentum included in the sum. This can clearly be seen in the quantum calculation at eq.~\eqref{eq:f_1_derivative}, and it can also be seen (as a naive product) in the definition of the heuristic picture of the energy momentum in eq.~\eqref{eq:heur_tensor_def}. Then the true sum over momenta should be reduced as $2\int^{\lambda_E \mathcal{T}/2}_0 dp\,p=(1/4)(\lambda_E \mathcal{T})^2$ in the heuristic picture, accounting for the difference.

\section{Conclusions and Extension to SU$(3)\times$U$(1)$}
\label{sec:conclusions}

The inheritance of angular momentum from background fields by means of the Schwinger effect to produced particles has been examined. Fields which were both though relevant to HIC and physically opaque were used; these were non-Abelian, SU$(2)$, fields which resemble Abelian projected fields in the color flux tube model~\cite{PhysRevD.20.179}, coupled with a homogeneous Abelian electromagnetic field with strong magnetic field component. The transport of angular momentum was reasoned through both a physically intuitive heuristic picture of pair production as well as an out-of-equilibrium calculation. It was found in both cases that the angular momentum transference was inhibited by a Schwinger exponential suppression, (i.e., $\exp(-\pi m^2/\lambda_E)$ with electric field $\lambda_E$ given by eq.~\eqref{eq:F_eigenvalues}), and moreover was proportional to the angular momentum of the gluonic background.

The heuristic picture of pair production stems from a virtual condensate breaking into particle-antiparticle pairs, whose trajectory initiated at arbitrary spacetime point follows classically according to Wong's equations, eq.~\eqref{eq:wong_lorentz}. Also, to rigorously confirm the transport of angular momentum from the background fields, a full quantum--to one-loop--out-of-equilibrium in-in calculation was performed. Both the heuristic picture and in-in formalism calculations were found to agree well with one-another. The mechanism for angular momentum transport in the heuristic picture is one of simple classical acceleration governed by Wong's equations.  However, we can confirm that the Schwinger effect is responsible for angular momentum transport through the in-in calculation. This is identifiable through the quadratic exponetial mass factor in the observable,  which only appears in the out-of-equilibrium contruction due to the Schwinger effect. 

To more carefully compare to the target environment of HICs, let us remark that our results also may be extensible to the case of SU$(3)\times$U$(1)$, and more generally to SU$(N)\times$U$(1)$. For relevance to HIC let us focus on SU$(3)$ though. The most prominent difference is whereas an isotropic color space exists for SU$(2)$, the color space of SU$(3)$ possesses a richer non-isotropic structure enabling background fields with color directional dependence. Let us follow refs.~\cite{PhysRevD.72.125010,TANJI20102018,PhysRevD.92.125012} for the SU$(3)$ color diagonalization. Consider a homogeneous field in SU$(3)$ such that for gauge field $\mathcal{A}_\mu^a(x) = \mathcal{A}_\mu(x) n^a$ for $n^a$ constant with color $a$. Then by means of a unitary transform $Un^a T^aU^{-1} = T^3 \cos \theta_c - T^8 \sin \theta_c =(1/2)\mathrm{diag}[\omega_1,\omega_2,\omega_3]\eqqcolon I_3$, where
\begin{equation}
    \omega_1 = \frac{2}{\sqrt{3}}\cos\bigl(\theta_c +\frac{\pi}{6}\bigr)\,,\quad
    \omega_2 = \frac{2}{\sqrt{3}}\cos\bigl(\theta_c +\frac{5\pi}{6}\bigr)\,,\quad
    \omega_3 = \frac{2}{\sqrt{3}}\cos\bigl(\theta_c +\frac{3\pi}{2}\bigr)\,.
\end{equation}
Here $T^a=(1/2)\lambda^a$ with $\lambda^a$ being the Gell-Mann matrices. 
$T^3$ and $T^8$ are the usual diagonal elements--one may find the eigenvalues of the field in terms of the second Casimir invariant of SU$(3)$, $C_2=(d^{abc}n^an^bn^c)^2 =(1/3)\sin^2 (3\theta_c)$
with $d^{abc}$ being the symmetric coefficients~\cite{Nayak:2005pf}, which project the color in a gauge in variant way. The salient point here is that for SU$(2)$ the couplings always come in equal and opposite pairs, however for SU$(3)$, the couplings need not be of the same magnitude, and a sum over effective couplings that may have color dependence is needed.

To extend our study from SU$(2)$ to SU$(3)$, one need only to replace the color isotropic sum over $\pm g$ to one over $\omega_n g$~\cite{TANJI20102018,PhysRevD.92.125012}. We have defined our non-Abelian fields in terms of isospin, eq.~\eqref{eq:homo_fields}, and for contrast with the SU$(2)$ case let us define our SU$(3)$ fields after diagonalization in the dressed propagator and kernel (eq.~\eqref{eq:K_pm} for SU$(2)$) with $\boldsymbol{\mathcal{E}}= \mathrm{tr}_c[I_3 G^{i0}]$ and likewise for the chromomagnetic field. Then for quantum calculations one need only sum over effective coupling $\omega_n g$ accompanied with the field strengths. Our key quantum observables in SU$(3)\times$U$(1)$ of the spin and angular momentum then can be found as
\begin{align}
    S^{0ij}&=-\sum_n \varepsilon^{kij}\frac{I_{\tilde{F}F}\mathcal{T}}{4\pi{}^{2}}\exp\Bigl(-\frac{\pi m^{2}}{\lambda_{E}}\Bigr)n_{\lambda_{E}}^{+\,k}\,,\\
   \boldsymbol{l}&\approx\sum_{n}\frac{\lambda_{B}}{(4\pi)^{2}}\frac{(\lambda_{E}\mathcal{T})^{2}}{\lambda_{E}^{2}+\lambda_{B}^{2}}e^{-\frac{\pi m^{2}}{\lambda_{E}}}\coth\Bigl(\frac{\pi\lambda_{B}}{\lambda_{E}}\Bigr)\boldsymbol{x}\times[\omega_{n}g\boldsymbol{\mathcal{E}}\times(e\boldsymbol{B}+\omega_{n}g\boldsymbol{\mathcal{B}})]\,.
   \label{eq:su3_ang_mom}
\end{align}
where now $I_{\tilde{F}F}=w_n g\boldsymbol{\mathcal{E}}\cdot(\omega_n g\boldsymbol{\mathcal{B}}+e\boldsymbol{B})$ and $I_{FF}=\omega_n^2 g^{2}\mathcal{B}^{2}+e^{2}B^{2}+2e\omega_ng\boldsymbol{B}\cdot\boldsymbol{\mathcal{B}}-\omega_n^2 g^{2}\mathcal{E}^{2}$. 

For simplicity let us look at two different cases of Abelian SU$(3)$ background fields, those proportional to $T^3$ (with, e.g., $\theta_c = \pi$) and those proportional to $T^8$, ($\pi/2$), and those with no preferential $\theta_c$ direction. The first case, $T^3= \mathrm{diag}[1/2,-1/2,0]$, provides an identical outcome as the SU$(2)$ $\sigma^3$ fields discussed throughout this paper, and therefore our results hold there. This is the scenario where the background field is only dependent on two colors, effectively decoupling one of the quarks~\cite{TANJI20102018}. Alternatively, for $T^8=(1/\sqrt{3})\mathrm{diag}[1/2,1/2,-1]$ fields with color field dependence there is a preferential likelihood that the same color anti-color quarks are produced in the pair production process. One of the most notable differences is that the conduction current, eq.~\eqref{eq:inin_current}, need not vanish~\cite{TANJI20102018}. In fact, for the case of a weak field and or a large mass, the $T^8$ conduction current would resemble an Abelian field. This would be the case for all observables with signature Schwinger pair production exponential suppression with weak fields/large mass. 

Finally, let us remark on the case relevant to HIC, and moreover the case as depicted in sec.~\ref{sec:heuristic}. We would find after averaging over events that still the orbital angular momentum is transferred by the Schwinger mechanism. Furthermore, let us assume an isotropy in color space such that there is no preferential $\theta_c$ direction, then we may average over $\theta_c$ in the final event average in SU$(3)\times$U$(1)$ to find that
\begin{equation}
   \frac{6}{\pi}\int_{0}^{\pi/6}d\theta\,\llangle\boldsymbol{l}\rrangle=
   \frac{2|\mathcal{B}_{\parallel}|}{3\sqrt{3}\pi^{3}}
   \coth\Bigl(\frac{\pi|\mathcal{B}_{\parallel}|}{|\mathcal{E}_{\parallel}|}\Bigr)\frac{\mathcal{E}_{\parallel}^{2}\mathcal{T}^{2}}{\mathcal{E}_{\parallel}^{2}+\mathcal{B}_{\parallel}^{2}}|g|^{3}\llangle\boldsymbol{l}_{\mathcal{F}}\rrangle\,.
\end{equation}
We have also assumed for the above calculation, a small mass such that the exponential suppression may be neglected. We find as anticipated in the SU$(2)\times$U$(1)$ that the orbital angular momentum is directly proportional to the background fields from which the pairs were created.

\section*{Acknowledgment}
We would like to thank Yoshitaka Hatta and Di-Lun Yang for valuable discussions. Y.H. was supported by JSPS KAKENHI Grant Numbers 17H06462 and 21H01084. P.C. would like to thank the Theory Center of KEK and YITP, Kyoto University where a portion of this work was accomplished.

\appendix

\section{Kernel Derivation}
\label{sec:kernel}

Here we derive the kernel given in Schwinger propertime, eq.~\eqref{eq:kernel}, in combinatory Abelian and non-Abelian SU(2) fields. The kernel is known to have an exact solution in homogeneous fields~\cite{PhysRev.82.664}. Our approach for its evaluation follows that used in ref.~\cite{doi:10.1142/S0217751X2030015X}. For the case of diagonal non-Abelian fields, i.e., $\propto \sigma_{3}$, we can express the kernel in an Abelian form, (we have made use of Wong's isospin equation solution as $I=(1/2)\sigma_{3}$), since the kernel may be decoupled as $\mathcal{K}(x,y,T)=\textrm{diag}(\mathcal{K}^{+}(x,y,T),\mathcal{K}^{-}(x,y,T))$ with
\begin{equation}
    \mathcal{K}^\pm(x,y,T)\coloneqq i\int\mathcal{D}x\mathcal{P}_{\gamma}e^{-i\int_{0}^{T}d\tau\bigl[m^{2}+\frac{1}{4}\dot{x}^{2}+eA_{\mu}\dot{x}^{\mu}\pm g\mathrm{tr}_c[I\mathcal{A}_{\mu}]\dot{x}^{\mu}+\frac{1}{2}(eF_{\mu\nu}\pm g \mathrm{tr}_c [IG_{\mu\nu}])\sigma^{\mu\nu}\bigr]}\,.
    \label{eq:K_pm}
\end{equation}
In the following we restrict our attention to the case of $\mathcal{K}^{+}(x,y,T)$, however, one can simply arrive at $\mathcal{K}^{-}(x,y,T)$ through the replacement $g\rightarrow -g$.

For homogeneous fields the kernel may be factored into a path integral
portion about fluctuating bosons as well as a spin factor portion,
let us begin with the latter which we write as
\begin{equation}
    \Phi(T)\coloneqq e^{-\frac{i}{2}\mathcal{F}_{\mu\nu}\sigma^{\mu\nu}T}\,.
\end{equation}
Then using the relationship $\frac{1}{2}\{\sigma^{\mu\nu},\sigma^{\alpha\beta}\}=g^{\mu\alpha}g^{\nu\beta}-g^{\nu\alpha}g^{\mu\beta}+i\gamma_{5}\varepsilon^{\mu\nu\alpha\beta}$, one can find that $(\mathcal{F}_{\mu\nu}\sigma^{\mu\nu})^{2}=4I_{FF}-8i\gamma_{5}I_{\tilde{F}F}$. And using the fact that $2^{2}(\lambda_{B}-i\gamma_{5}\textrm{sgn}(I_{\tilde{F}F})\lambda_{E})^{2}=4I_{FF}-8i\gamma_{5}I_{\tilde{F}F}$ as well one can find that
\begin{equation}
    \sin\Bigl(\frac{1}{2}\mathcal{F}_{\mu\nu}\sigma^{\mu\nu}T\Bigr)=\frac{1}{2}\bigl(\lambda_{B}+i\gamma_{5}\textrm{sgn}(I_{\tilde{F}F})\lambda_{E}\bigr)\frac{\mathcal{F}_{\mu\nu}\sigma^{\mu\nu}}{\lambda_{B}^{2}+\lambda_{E}^{2}}\sin\bigl[(\lambda_{B}-i\gamma_{5}\textrm{sgn}(I_{\tilde{F}F})\lambda_{E})T\bigr]\,.
\end{equation}
Then the spin factor becomes
\begin{equation}
    \Phi(T)=\cos\bigl[(\lambda_{B}-i\gamma_{5}\textrm{sgn}(I_{\tilde{F}F})\lambda_{E})T\bigr]-i\sin\Bigl(\frac{1}{2}\mathcal{F}_{\mu\nu}\sigma^{\mu\nu}T\Bigr)\,.
\end{equation}
Noting that
\begin{align}
    \sin\bigl[(\lambda_{B}-i\gamma_{5}\textrm{sgn}(I_{\tilde{F}F})\lambda_{E})T\bigr]
    &=\sin(\lambda_BT) \cosh(\lambda_ET) -i\gamma_5\textrm{sgn}(I_{\tilde{F}F})\cos(\lambda_BT)\sinh(\lambda_ET), \\
    \cos\bigl[(\lambda_{B}-i\gamma_{5}\textrm{sgn}(I_{\tilde{F}F})\lambda_{E})T\bigr]
    &=\cos(\lambda_BT) \cosh(\lambda_ET)+i\gamma_5\textrm{sgn}(I_{\tilde{F}F})\sin(\lambda_BT)\sinh(\lambda_ET),
\end{align}
one may finally write the spin factor as
\begin{align}
    \Phi(T)&=\cos(\lambda_{B}T)\cosh(\lambda_{E}T)+i\gamma_{5}\textrm{sgn}(I_{\tilde{F}F})\sin(\lambda_{B}T)\sinh(\lambda_{E}T) -\frac{1}{2}\bigl[\lambda_{B}+i\gamma_{5}\textrm{sgn}(I_{\tilde{F}F})\lambda_{E}\bigr] \nonumber\\
    &\times\frac{\mathcal{F}_{\mu\nu}\sigma^{\mu\nu}}{\lambda_{B}^{2}+\lambda_{E}^{2}}\Bigl[i\sin(\lambda_{B}T)\cosh(\lambda_{E}T)+\gamma_{5}\textrm{sgn}(I_{\tilde{F}F})\cos(\lambda_{B}T)\sinh(\lambda_{E}T)\Bigr]\,.
    \label{eq:kernel_spin}
\end{align}

Let us now address the bosonic path integral portion. The path integral in $\mathcal{K}^{+}$, eq.~\eqref{eq:K_pm}, is
\begin{equation}
    b(x,y,T)= \int\mathcal{D}x\,e^{iS_{\textrm{B}}}\,,\quad S_{\textrm{B}}= -\int_{0}^{T}d\tau\Bigl[\frac{1}{4}\dot{x}^{2}+eA_{\mu}\dot{x}^{\mu}+g\mathrm{tr}_c(I\mathcal{A}_{\mu})\dot{x}^{\mu}\Bigr]\,.
\end{equation}
Since the action is quadratic in $x$ the path integral may be evaluated exactly. We expand about $x(\tau)=x_{cl}+\eta(\tau)$, with the classical path obeying $\ddot{x}_{cl}(\tau)=2\mathcal{F}\dot{x}_{cl}(\tau)$, c.f., eq.~\eqref{eq:wong_lorentz}. Then we have for Dirichlet boundary conditions, i.e., $\eta(0)=\eta(T)=0$,
\begin{equation}
    b(x,y,T)=e^{iS_{\textrm{B}}(x^{cl})}\mathcal{A}_{fl}\,,\quad \mathcal{A}_{fl}= \int\mathcal{D}\eta\exp\Bigl\{ i\int_{0}^{T}d\tau\Bigl[-\frac{1}{4}\dot{\eta}^{2}+\frac{1}{2}\eta^{\mu}\mathcal{F}_{\mu\nu}\dot{\eta}^{\nu}\Bigr]\Bigr\}\,.
\end{equation}

Let us first calculate the classical worldline action. Since $x_{cl}(T)=x$ and $x_{cl}(0)=y$, one can find for $\dot{x}(\tau)=\exp(2\mathcal{F}\tau)\dot{x}(0)$ the following: 
\begin{equation}
    z\coloneqq x-y=\frac{1}{2}\mathcal{F}^{-1}(e^{2\mathcal{F}T}-1)\dot{x}_{cl}(0)\,,
\end{equation}
from which it follows that $\dot{x}_{cl}(T)+\dot{x}_{cl}(0)=2\coth(\mathcal{F}T)\mathcal{F}z$ and $\dot{x}_{cl}(T)-\dot{x}_{cl}(0)=2\mathcal{F}z$. Next let us select a gauge; we use the Fock-Schwinger gauge: $eA(x_{cl})+g \mathrm{tr}_c[I\mathcal{A}(x_{cl})]=-(1/2)\mathcal{F}x_{cl}$.
Then one can find for the gauge dependent action
\begin{equation}
    \varphi(x,y,T)\coloneqq S_{\textrm{B}}(x^{cl})=\frac{1}{2}x\mathcal{F}y-\frac{1}{4}z^T\coth(\mathcal{F}T)\mathcal{F}z\,.
\end{equation}
Making use of the relations $\cosh(\mathcal{F}T) =\cosh(\lambda_{E}T)P_{E}+\cos(\lambda_{B}T)P_{B}$ and $\sinh(\mathcal{F}T)=\lambda_{E}^{-1}\mathcal{F}\sinh(\lambda_{E}T)P_{E}+\lambda_{B}^{-1}\mathcal{F}\sin(\lambda_{B}T)P_{B}$ where the projection operators are given in eq.~\eqref{eq:proj_op}, one can express the above as
\begin{equation}
    \varphi(x,y,T)=\frac{1}{2}x^T\mathcal{F}y-\frac{1}{4}\Bigl\{\lambda_{E}\coth(\lambda_{E}T)z^TP_{E}z+\lambda_{B}\cot(\lambda_{B}T)z^TP_{B}z\Bigr\}\,.
    \label{eq:kernel_action}
\end{equation}

The last step is to calculate the fluctuation prefactor which owing to its quadratic form can be written as
\begin{equation}
    \mathcal{A}_{fl}=\det\Bigl[\frac{1}{4T}\frac{d^{2}}{d\tau^{2}}+\frac{1}{2}\mathcal{F}\frac{d}{d\tau}\Bigr]^{-\frac{1}{2}}\det\Bigl[\frac{1}{4T}\frac{d^{2}}{d\tau^{2}}\Bigr]^{\frac{1}{2}}\bigg/\int\mathcal{D}\eta\exp\Bigl\{-i\int_{0}^{1}d\tau\frac{1}{4T}\dot{\eta}^{2}\Bigr\}\,.
\end{equation}
In this expression, we rescaled $\tau \to T\tau$ such that the period of $\tau$ is from $0$ to $1$.
To complete the functional determinant, note that since the field strength tensor is independent of time, we can find a similarity transform such that for eigenvalues of $\mathcal{F}$ we have 
\begin{equation}
    \det\Bigl[\frac{1}{4T}\frac{d^{2}}{d\tau^{2}}+\frac{1}{2}\mathcal{F}\frac{d}{d\tau}\Bigr] =\det\Bigl[\frac{1}{4T}\frac{d^{2}}{d\tau^{2}}+\frac{1}{2}D_\mathcal{F}\frac{d}{d\tau}\Bigr]\,,
\end{equation}
where $D_\mathcal{F}=\textrm{diag}(\lambda_{E},-\lambda_{E},i\lambda_{B},-i\lambda_{B})$. Let us look at $\lambda_{E}$. Since we now have one-dimensional operators with Dirichlet boundary conditions we evaluate the determinants through the Gel'fand Yaglom technique~\cite{doi:10.1063/1.1703636,10.2307/2041911}. Then for $[(4T)^{-1}d^2/d\tau^2+(\lambda_{E}/2)d/d\tau]u_{\lambda_E} =0$ one can find $u_{\lambda_E} =(2\lambda_{E}T)^{-1}[1-\exp(-2\lambda_{E}T\tau)]$ that satisfies $u_{\lambda_E}(0)=0$ and $\dot{u}_{\lambda_E}(0)=1$ and similarly for the other eigenvalues. For the case of zero eigenvalue, $[(4T)^{-1}d^2/d\tau^2]u_0=0$, we find $u_0(1)=1$. The determinant can then be found as
\begin{equation}
    \frac{\det\bigl[\frac{1}{4T}\frac{d^{2}}{d\tau^{2}}+\frac{1}{2}\mathcal{F}\frac{d}{d\tau}\bigr]}{\det\bigl[\frac{1}{4T}\frac{d^{2}}{d\tau^{2}}\bigr]}=\frac{u_{\lambda_E}(1)u_{-\lambda_E}(1)u_{i\lambda_B}(1)u_{-i\lambda_B}(1)}{u_{0}(1)^{4}}=\frac{\sinh^{2}(\lambda_{E}T)\sin^{2}(\lambda_{B}T)}{\lambda_{E}^{2}\lambda_{B}^{2}T{}^{4}}\,.
\end{equation}
And the normalizing factor can be found as $\int\mathcal{D}\eta\,\exp[-i\int_{0}^{1}d\tau(4T)^{-1}\dot{\eta}^{2}] =-i(4\pi T)^{-2}$. Finally we can gather all of the pieces of the kernel to find
\begin{equation}
    \mathcal{K}^{+}(x,y,T)=\frac{\lambda_{E}\lambda_{B}\exp\bigl[-im^{2}T+i\varphi(x,y,T)\bigr]}{(4\pi)^{2}\sinh(\lambda_{E}T)\sin(\lambda_{B}T)}\Phi(T)\,.
\end{equation}

\bibliographystyle{ptephy}
\bibliography{references}

\end{document}